\documentclass[aps,prd,twocolumn,twoside,superscriptaddress,floatfix, nofootinbib]{revtex4-1}
\usepackage[T1]{fontenc}

\providecommand{\sorthelp}[1]{}
 
\usepackage{graphicx}
\usepackage{amssymb}
\usepackage{amsmath}
\usepackage{multirow}
\usepackage[colorlinks=true,linkcolor=blue,citecolor=blue]{hyperref}%
\usepackage{graphicx}
\usepackage{bm}
\usepackage{xcolor}

\usepackage{orcidlink}

\newcommand{\beq} {\begin{equation}}
\newcommand{\eeq} {\end{equation}}
\newcommand{\bal} {\begin{aligned}}
\newcommand{\eal} {\end{aligned}}
\newcommand*\Bell{\ensuremath{\boldsymbol\ell}}

\usepackage[caption=false]{subfig}
\usepackage[capitalise]{cleveref}
\usepackage{mhchem}

\newcommand{\bl}{\boldsymbol{\ell}}
\newcommand{\be}{\begin{eqnarray}}

\newcommand{\ee}{\end{eqnarray}}

\usepackage[normalem]{ulem}

\usepackage[export]{adjustbox}

\providecommand{\sorthelp}[1]{}

\begin{document}
\title{The Atacama Cosmology Telescope: A search for late-time anisotropic screening of the Cosmic Microwave Background}

\newcommand{\cca}{CCA}
\author{William~R.~Coulton}
\affiliation{Kavli Institute for Cosmology Cambridge, Madingley Road, Cambridge CB3 0HA, UK}
\affiliation{DAMTP, Centre for Mathematical Sciences, University of Cambridge, Cambridge CB3 OWA, UK}
\author{Theo Schutt}
\affiliation{ SLAC National Accelerator Laboratory, Menlo Park, California 94025, USA}
\affiliation{Kavli Institute for Particle Astrophysics and Cosmology, Stanford, CA  94305-4060, USA}
\affiliation{Department of Physics, Stanford University, Stanford, CA 94305, USA}
\author{ Abhishek S. Maniyar}
\affiliation{ SLAC National Accelerator Laboratory, Menlo Park, California 94025, USA}
\affiliation{Kavli Institute for Particle Astrophysics and Cosmology, Stanford, CA  94305-4060, USA}
\affiliation{Department of Physics, Stanford University, Stanford, CA 94305, USA}
\author{ Emmanuel~Schaan}
\affiliation{ SLAC National Accelerator Laboratory, Menlo Park, California 94025, USA}
\affiliation{Kavli Institute for Particle Astrophysics and Cosmology, Stanford, CA  94305-4060, USA}
\affiliation{Department of Physics, Stanford University, Stanford, CA 94305, USA}
\author{ Rui An}
\affiliation{Department of Physics and Astronomy, University of Southern California, Los Angeles, CA 90089, USA}
\author{ Zachary~Atkins}
\affiliation{Joseph Henry Laboratories of Physics, Jadwin Hall, Princeton University, Princeton, NJ, USA 08544}
\author{ Nicholas Battaglia}
\affiliation{ Department of Astronomy, Cornell University, Ithaca, NY 14853, USA}
\author{ J~Richard~Bond}
\affiliation{Canadian Institute for Theoretical Astrophysics, University of Toronto, Toronto, ON, Canada M5S 3H8}
\author{ Erminia Calabrese}
\affiliation{School of Physics and Astronomy, Cardiff University, Cardiff, Wales CF24 3AA, UK}
\author{ Steve~K.~Choi}
\affiliation{Department of Physics and Astronomy, University of California, Riverside, CA 92521, USA}
\affiliation{ Department of Astronomy, Cornell University, Ithaca, NY 14853, USA}
\author{ Mark~J.~Devlin}
\affiliation{Department of Physics and Astronomy, University of Pennsylvania, Philadelphia, PA, USA 19104}
\author{ Adriaan J. Duivenvoorden}
\affiliation{ Center for Computational Astrophysics, Flatiron Institute, New York, NY 10010 USA}
\affiliation{Joseph Henry Laboratories of Physics, Jadwin Hall, Princeton University, Princeton, NJ, USA 08544}
\author{ Jo~Dunkley}
\affiliation{Joseph Henry Laboratories of Physics, Jadwin Hall, Princeton University, Princeton, NJ, USA 08544}
\affiliation{Department of Astrophysical Sciences, Peyton Hall, Princeton University, Princeton, NJ USA 08544}
\author{ Simone~Ferraro}
\affiliation{Physics Division, Lawrence Berkeley National Laboratory, Berkeley, CA, USA}
\affiliation{Department of Physics, University of California, Berkeley, CA, USA 94720}
\author{ Vera Gluscevic}
\affiliation{Department of Physics and Astronomy, University of Southern California, Los Angeles, CA 90089, USA}
\author{ J.~Colin~Hill}
\affiliation{Department of Physics, Columbia University, New York, NY, USA}
\affiliation{ Center for Computational Astrophysics, Flatiron Institute, New York, NY 10010 USA}
\author{ Matt~Hilton}
\affiliation{Wits Centre for Astrophysics, School of Physics, University of the Witwatersrand}
\affiliation{School of Mathematics, Statistics \& Computer Science, University of KwaZulu-Natal}
\author{ Adam D. Hincks}
\affiliation{David A. Dunlap Department of Astronomy \& Astrophysics, University of Toronto}
\affiliation{Specola Vaticana (Vatican Observatory), V-00120 Vatican City State}
\author{ Arthur Kosowsky}
\affiliation{Department of Physics and Astronomy, University of Pittsburgh, Pittsburgh PA 15260 USA}
\author{ Darby~Kramer}
\affiliation{School of Earth and Space Exploration, Arizona State University, Tempe, AZ, USA 85287}
\author{  Aleksandra~Kusiak}
\affiliation{Department of Physics, Columbia University, New York, NY, USA}
\author{  Adrien La Posta}
\affiliation{University of Oxford, Denys Wilkinson Building, Keble Road, Oxford, OX1 3RH, UK}
\author{  Thibaut~Louis}
\affiliation{ Universit\'e Paris-Saclay, CNRS/IN2P3, IJCLab, 91405 Orsay, France}
\author{ Mathew~S.~Madhavacheril}
\affiliation{Department of Physics and Astronomy, University of Pennsylvania, Philadelphia, PA, USA 19104}
\author{ Gabriela A. Marques}
\affiliation{Centro Brasileiro de Pesquisas Físicas,  R. Dr. Xavier Sigaud, 150 - Botafogo, Rio de Janeiro - RJ, 22290-180}
\affiliation{Fermi National Accelerator Laboratory, Batavia, IL 60510, USA}
\author{ Fiona~McCarthy}
\affiliation{DAMTP, Centre for Mathematical Sciences, University of Cambridge, Cambridge CB3 OWA, UK}
\affiliation{ Center for Computational Astrophysics, Flatiron Institute, New York, NY 10010 USA}
\author{ Jeff~McMahon}
\affiliation{Kavli Institute for Cosmological Physics, University of Chicago, Chicago, IL 60637, USA}
\affiliation{Department of Astronomy and Astrophysics, University of Chicago, Chicago, IL 60637, USA}
\affiliation{Department of Physics, University of Chicago, Chicago, IL 60637, USA}
\affiliation{Enrico Fermi Institute, University of Chicago, Chicago, IL 60637, USA}
\author{ Kavilan~Moodley}
\affiliation{ Astrophysics Research Centre, University of KwaZulu-Natal, Westville Campus,Durban 4041, South Africa}
\affiliation{School of Mathematics, Statistics \& Computer Science, University of KwaZulu-Natal}
\author{ Sigurd~Naess}
\affiliation{Institute of Theoretical Astrophysics, University of Oslo, Norway}
\author{ Lyman~A.~Page}
\affiliation{Joseph Henry Laboratories of Physics, Jadwin Hall, Princeton University, Princeton, NJ, USA 08544}
\author{ Bruce~Partridge}
\affiliation{Department of Physics and Astronomy, Haverford College, Haverford, PA, USA 19041}
\author{ Frank~J.~Qu}
\affiliation{DAMTP, Centre for Mathematical Sciences, University of Cambridge, Cambridge CB3 OWA, UK}
\affiliation{Kavli Institute for Cosmology Cambridge, Madingley Road, Cambridge CB3 0HA, UK}
\author{ Neelima Sehgal}
\affiliation{Physics and Astronomy Department, Stony Brook University, Stony Brook, NY 11794}
\author{ Blake D.~Sherwin}
\affiliation{DAMTP, Centre for Mathematical Sciences, University of Cambridge, Cambridge CB3 OWA, UK}
\affiliation{Kavli Institute for Cosmology Cambridge, Madingley Road, Cambridge CB3 0HA, UK}
\author{ Crist\'obal~Sif\'on}
\affiliation{Instituto de F{\'{i}}sica, Pontificia Universidad Cat{\'{o}}lica de Valpara{\'{i}}so, Valpara{\'{i}}so, Chile}
\author{ David~N.~Spergel}
\affiliation{ Center for Computational Astrophysics, Flatiron Institute, New York, NY 10010 USA}
\affiliation{Department of Astrophysical Sciences, Peyton Hall, Princeton University, Princeton, NJ USA 08544}
\author{ Suzanne~T.~Staggs}
\affiliation{Joseph Henry Laboratories of Physics, Jadwin Hall, Princeton University, Princeton, NJ, USA 08544}
\author{ Alexander~Van~Engelen}
\affiliation{School of Earth and Space Exploration, Arizona State University, Tempe, AZ, USA 85287}
\author{ Cristian~Vargas}
\affiliation{Instituto de Astrof\'isica and Centro de Astro-Ingenier\'ia, Facultad de F\`isica}
\affiliation{Pontificia Universidad Cat\'olica de Chile, Santiago, Chile}
\author{  Edward~J.~Wollack}
\affiliation{NASA/Goddard Space Flight Center, Greenbelt, MD, USA 20771}

\begin{abstract}
    
Since the formation of the first stars, most of the gas in the Universe has been ionized.
Spatial variations in the density of this ionized gas generate cosmic microwave background anisotropies via Thomson scattering, a process known as the ``anisotropic screening'' effect. 
We propose and implement for the first time a new estimator to cross-correlate \textit{unWISE} galaxies and anisotropic screening, as measured by the Atacama Cosmology Telescope and \textit{Planck} satellite. 
We do not significantly detect the effect; the null hypothesis is consistent with the data 
at 1.7~$\sigma$ (resp. 0.016~$\sigma$) for the blue (resp. green) unWISE sample.
We obtain an upper limit on the integrated optical depth within a 6 arcmin disk  to be $\bar{\tau}< 0.033$ arcmin$^2$ at 95\% confidence for the blue sample and $\bar{\tau}< 0.057$ arcmin$^2$ for the green sample.
Future measurements with Simons Observatory and CMB-S4 should detect this effect significantly.
Complementary to the kinematic Sunyaev-Zel'dovich effect, this probe of the gas distribution around halos will inform models of feedback in galaxy formation and baryonic effects in galaxy lensing.
\end{abstract}

\maketitle
\section{Introduction}
\label{sec:intro}
The distribution of gas throughout the Universe contains a wealth of information on the astrophysical processes governing the environments around galaxies and galaxy clusters \cite{Ostriker_2005,Nagai_2007,McNamara_2007,Battaglia_2010,Moser_2022}.
Whilst the properties of gas in the center of clusters have been well characterized \cite{McNamara_2007,Ruan_2015,Crichton_2016,deGraaff_2019,Eckert_2019,Zenteno_2020,Pandey_2022}, the gas distribution in cluster outskirts and surrounding lower mass systems (the circumgalactic medium (CGM)) is poorly understood and is the subject of many recent works \citep[e.g.][]{planck2012-VIII,Bonjean_2018,Tanimura_2019,deGraaff_2019,Hincks_2022,Isopi_2024}.
Processes like supernovae and jets from supermassive black holes can eject material into the CGM, strongly influencing galaxy formation, but the details of these processes are not well constrained \cite{Tumlinson_2017}.
Thus, measurements of the CGM can inform galaxy formation and evolution models and state-of-the-art hydrodynamical simulations \cite{Moser_2022}.
Further, since these feedback processes redistribute the baryonic matter of galaxies, modeling their effect on estimates of galactic mass is a key ingredient of cosmological analyses, especially those using the weak gravitational lensing of distant galaxies, e.g., Refs. \cite{Amon_2022,Arico_2023}.

Measuring the CGM is difficult, as it emits no optical or infrared light, and it is too cool ($\sim 10^4$\,K) to radiate effectively in X-rays. However, several methods exist for probing the CGM that use the Thomson scattering of cosmic microwave background (CMB) photons by this diffuse ionized gas.

The CMB is the remnant light from the very early Universe, produced $\sim$400,000 years after the Big Bang, observed as a two-dimensional surface across the sky.
The small fluctuations in matter density at this epoch created the degree-scale fluctuations, or ``primary anisotropies'', observed in the CMB temperature today \cite{planck2016-l01}.
As these photons traverse space, they can also interact with the CGM, leaving a ``shadow'' of the gas on the CMB.

Two shadow sources are the thermal and kinetic Sunyaev-Zel'dovich (tSZ and kSZ) effects \cite{Sunyaev_1972,Sunyaev_1980}.
The tSZ effect is due to the change in energy of CMB photons scattered by free electrons due to the electrons' thermal velocities and thus is proportional to the gas \emph{pressure} integrated along the line of sight (LOS).
The kSZ effect is due to the Doppler shift in the photons' energy due to the bulk velocity of the gas along the LOS ($v_{\rm LOS}$) and is proportional to the integrated line-of-sight product of $v_{\rm LOS}$ and the gas \emph{density}.
Whilst these effects have been used to measure the thermodynamic properties of the ionized gas around galaxies \citep{Amodeo_2021,Liu_2025,Hadzhiyska_2025a}, new measurements are needed to understand the physics governing the gas properties.

A third, currently undetected scattering effect that can directly probe the density of ionized gas is the ``anisotropic screening'' effect.
Screening occurs when CMB photons along the LOS are Thomson scattered out of the LOS. 
With the same probability, CMB photons from every other direction are scattered into the LOS.  
On average the photons scattered into the LOS have the average temperature of the CMB, and thus screening slightly damps the primary anisotropies. 
The isotropic component scales with the optical depth through reionization and is one of the six standard cosmological parameters \cite{planck2014-a10,deBelsunce_2021}.
However, the damping also has an anisotropic component where lines of sight with more (fewer) electrons will be more (less) suppressed. 
This spatially modulates the underlying CMB anisotropies \cite{Ostriker_1986,Vishniac_1987,Dvorkin_2009}. 
This effect is known as the ``anisotropic screening'' effect and, on small scales ($\theta \lesssim 1^{\circ}$) and for the expected level of scattering, the new anisotropies are
\begin{align}\label{eq:screeningEffect}
\Delta T^\text{anisotropic screen}(\mathbf{\hat{n}}) = -\delta\tau(\mathbf{\hat{n}})\Delta T^\mathrm{primary}(\mathbf{\hat{n}}),
\end{align}
where $\delta \tau (\mathbf{\hat{n}})$ is the fluctuation in the optical depth to
last scattering in direction $\mathbf{\hat{n}}$ 
and $\Delta T_{\rm primary}(\mathbf{\hat{n}})$ is
the primordial CMB anisotropy in that direction.

Measuring anisotropic screening is thus a tantalizing prospect: it provides a direct probe of the gas density complementary to tSZ and kSZ, but the induced change in CMB temperature has the smallest magnitude among the scattering effects. 
This paper describes a search for the anisotropic screening effect through the cross correlation of CMB maps with optical galaxy catalogs. 
We implement a new method to cleanly disentangle it from contaminant signals. 
In \cref{sec:stacking} we describe our methodology and a set of validation tests in \cref{sec:validation}. In \cref{sec:data} we introduce our data sets and then discuss our results in \cref{sec:results}. We present our conclusions in \cref{sec:conclusion}.

\section{Cross-Correlation Methodology }\label{sec:stacking}

A key challenge to detecting anisotropic screening is disentangling it from the host of other small-scale sky signals, such as emission from high-redshift star-forming galaxies and radio galaxies, gravitational lensing, and the thermal and kinetic Sunyaev-Zeldovich effects (tSZ \& kSZ) see, e.g., \cite{Aghanim_2008}. 
The foreground anisotropies are expected to be at least 30 times larger than the anisotropic screening signal and are spatially correlated with it. 
To avoid potential biases, we combine well-tested multifrequency component-separation methods with a new estimator called the ``sign estimator'' \cref{eq:stackedEst}.

In this Section we first review the anisotropic screening effect, then the standard estimator as derived in Ref.~\cite{Dvorkin_2009} before presenting our new method.  We refer the reader to Ref.~\cite{Schutt_2023} for a detailed analysis of the relative merits of the different estimators.

\subsection{The anisotropic screening effect}

In the screening effect, free electrons along a given LOS Thomson scatter a fraction $1-e^{-\tau}$ of the CMB photons from this LOS away from the observer \cite{Hu_1994,Dodelson_1995,Persi_1995,Gruzinov_1998,Dvorkin_2009a}.
At the same time, Thomson scattering by the same free electrons deflects other CMB photons (which would not otherwise reach the observer) into the observer's LOS.
Because Thomson scattering is a time-reversible process, the probability for a LOS photon to be deflected out of the LOS is the same as the probability for any photon to be deflected into the LOS.
Thus, screening is simply replacing a fraction $1-e^{-\tau}$ of LOS photon flux at $\tilde{T}(\mathbf{\hat{n}})$ with an equal fraction $1-e^{-\tau}$ of all other photons incident on the electron, whose average temperature is $\bar{T}_\text{CMB}$\cite{Dvorkin_2009}:
\beq
T(\mathbf{\hat{n}})
=
\tilde{T}(\mathbf{\hat{n}}) 
e^{-\tau}
+
\bar{T}_\text{CMB}
\left[ 1 - e^{-\tau(\mathbf{\hat{n}})}  \right],
\eeq
where $T(\mathbf{\hat{n}})$ is the observed temperature anisotropy and $\tilde{T}(\mathbf{\hat{n}})$ is the CMB anisotropies in the absence of screening.
Note that we have dropped the quadrupole scattering term as this is negligible on small scales. 
This gives the observed temperature fluctuation:
\beq
\bal
&\Delta T(\mathbf{\hat{n}})
\equiv
T(\mathbf{\hat{n}})
-
\bar{T}_\text{CMB}\\
&=
\Delta \tilde{T}(\mathbf{\hat{n}}) 
e^{-\tau(\mathbf{\hat{n}})}
\quad \text{with  }
\Delta \tilde{T}(\mathbf{\hat{n}}) \equiv \tilde{T}(\mathbf{\hat{n}}) - \bar{T}_\text{CMB},\\
&=
\Delta \tilde{T}(\mathbf{\hat{n}}) 
e^{-\bar{\tau}}
e^{-\delta\tau(\mathbf{\hat{n}})}
\quad \text{with  }
\tau(\mathbf{\hat{n}}) = \bar{\tau} + \delta\tau(\mathbf{\hat{n}}),\\
&=
\Delta T^\text{primary}(\mathbf{\hat{n}})
e^{-\delta\tau(\mathbf{\hat{n}})}
\quad \text{with  }
\Delta T^\text{primary}(\mathbf{\hat{n}}) \equiv \Delta \tilde{T}(\mathbf{\hat{n}}) 
e^{-\bar{\tau}},\\
&\simeq
\Delta T^\text{primary}(\mathbf{\hat{n}})
\left[ 1 - \delta\tau(\mathbf{\hat{n}}) \right].
\eal
\eeq
Here we have introduced the average optical depth $\bar{\tau}$ to Thomson scattering, through which the unscreened fluctuations $\Delta \tilde{T}(\mathbf{\hat{n}})$ are observed as $\Delta T^\text{primary}(\mathbf{\hat{n}})$.
While this mean optical depth is not negligible ($\bar{\tau} \sim 5-6\%$ from Planck \cite{planck2014-a10,Pagano_2020,planck2020-LVII,deBelsunce_2021}) the fluctuations $\delta\tau(\mathbf{\hat{n}})$ in the optical depth are much smaller, of order $10^{-5}-10^{-3}$ for galaxy/cluster halos, justifying the linear expansion in the last line above. 
Anisotropic screening thus in principle adds power to the primary CMB power spectrum on small scales. However, this contribution is dominated by the kSZ effect, and thus it is very difficult to measure anisotropic screening directly via the CMB power spectrum.

Screening of the CMB also occurs in polarization. 
The equations above remain exact when replacing the temperature $T$ by the Stokes parameters $Q$ and $U$.
In terms of $E$ and $B$ instead, the multiplication of $Q$ and $U$ with a scalar function $1-\delta\tau(\mathbf{\hat{n}})$ mixes $E$ and $B$ modes, thus leading to the generation of $B$-modes from $E$-modes \cite{Mortonson_2007,Dore_2007,Dvorkin_2009,Dvorkin_2009a,Su_2011,Roy_2018,Roy_2021}.
An additional source of polarization is the conversion of the primary temperature quadrupole to linear polarization (E and B modes). The screening induced polarization anisotropies are smaller than the temperature anisotropies and so are not considered in this work.

\subsection{The quadratic estimator}
\label{sec:quadraticEst}

As described in Ref. \cite{Dvorkin_2009}, we can construct a quadratic estimator for $\delta\tau$ as\footnote{In our notation $\int_{\bl} \equiv \int\frac{\mathrm{d}^2\bl}{\left(2\pi\right)^2}$}
\begin{align}
    &\delta\tilde{\tau}(\mathbf{L}) =\int_{\bl_1\bl_2} 
    g(\bl_1,\bl_2)T(\bl_1)T(\bl_2) (2\pi)^2\delta^{(2)}(\mathbf{L}-\bl_1-\bl_2)
\end{align}
where $T(\bl)$ are the 2D Fourier modes of the observed CMB temperature fluctuation map, as the intuition is clearer in the flat-sky approximation, $g(\bl_1,\bl_2)$ are a set of weights and $\delta^{(2)}$ is the 2D Dirac delta function. 
The weights are found by searching for the minimum variance unbiased estimator giving
\begin{align}\label{eq:optimal_qe}
    &\delta\tilde{\tau}(\mathbf{L}) =-\frac{1}{N_L}\int_{\bl_1\bl_2} 
C^{TT}_{\bl_1}\frac{T(\bl_1)T(\bl_2)}{C^\mathrm{tot}_{\bl_1}C^\mathrm{tot}_{\bl_2} }(2\pi)^2\delta^{(2)}(\mathbf{L}-\bl_1-\bl_2)
\end{align}
where $C^{TT}_\ell$ is the power spectrum of the $  \tilde{T}(\mathbf{\hat{n}})e^{-\bar{\tau}}$ field, i.e. the standard CMB power spectrum, $C^\mathrm{tot}_\ell $ is the total power spectrum including foregrounds and instrumental noise and $N_L$ is a normalization constant given by
\begin{align}
N_L = 2\int _{\bl_1\bl_2}
(2\pi)^2\delta^{(2)}(\mathbf{L}-\bl_1-\bl_2) \frac{\left.C^{TT}_{\bl_1}\right.^2}{C^\mathrm{tot}_{\bl_1}C^\mathrm{tot}_{\bl_2} }. 
\end{align}

\subsection{General small-scale $\tau$ estimator}\label{sec:smallScales}

For small scale reconstructions, $L\gtrsim$ 2000, on data sets with noise and foregrounds, \cref{eq:optimal_qe} can be approximated as 
\begin{align}\label{eq:smallScaleQuadEstimator}
    \delta\tilde{\tau}(\mathbf{L}) \approx &-\int \mathrm{d}^2\mathbf{\hat{n}} e^{i\mathbf{\hat{n}}\cdot\mathbf{L}}  T_\mathrm{large-scales}(\mathbf{\hat{n}})  \left[{C^\mathrm{tot}}^{-1} T_\mathrm{small-scales}\right](\mathbf{\hat{n}}) \nonumber \\ &\times
    C^\mathrm{tot}_L\left[\int_{\ell_\mathrm{small-scales}} \frac{\mathrm{d}^2\bl}{(2\pi)^2}C^{TT}_{\ell} \right]^{-1},
\end{align}
where $ T_\mathrm{large-scales}(\mathbf{\hat{n}})$ is a map that contains only the large-scale modes of the map, $\left[{C^\mathrm{tot}}^{-1}  T_\mathrm{small-scales}\right](\mathbf{\hat{n}}) $ is an inverse variance filtered map of small-scale modes and $\ell_\mathrm{small-scales}$ are the modes that contribute to the small-scale map. This estimator arises from  three parts. First, consider the Wiener filter $C^{TT}_{\ell_1}/C^\mathrm{tot}_{\ell_1}$ in \cref{eq:optimal_qe}:
this is unity on large-scales where the CMB dominates and zero on small scales where noise and foregrounds dominate. 
Thus, this Wiener filter operation works like an effective low-pass filter to make a map of the large-scale primary CMB anisotropies. Second, consider the Dirac delta function in \cref{eq:optimal_qe}. We are reconstructing small scales, i.e., large $\mathbf{L}$, and $\mathbf{\ell}_1$ is constrained to be small due to the $C^{TT}/C^\mathrm{tot}$ filtering. Thus, the Dirac delta function forces $\mathbf{\ell}_2\sim \mathbf{L}$, so the second part of the estimator selects small-scale modes. By writing the Dirac delta function as
\begin{align}
    (2\pi)^2\delta^{(2)}(\mathbf{L}-\Bell_1-\Bell_2) = \int \mathrm{d}n^2 e^{i\mathbf{n}\cdot(\mathbf{L}-\Bell_1-\Bell_2) },
\end{align} we express the small-scale and large-scale terms as real space maps.
Finally, we have approximated $C^{tot}_{\bl+\mathbf{L}}\approx C^{tot}_{\mathbf{L}}$, which is valid on small scales where the power spectrum varies slowly. The geometric constraint, from the Dirac delta function, allows one final simplification: the cancellation of the inverse variance filter and the $C^\mathrm{tot}_L$ factors to give

\begin{align}\label{eq:smallScaleQuadEstimatorReduced}
    \delta\tilde{\tau}(\mathbf{L}) \approx &-\int \mathrm{d}^2\mathbf{\hat{n}} e^{i\mathbf{\hat{n}}\cdot\mathbf{L}}  T_\mathrm{large-scales}(\mathbf{\hat{n}}) T_\mathrm{small-scales}(\mathbf{\hat{n}}) \nonumber \\ &\times
    \left[\int_{\ell_\mathrm{small-scales}} \frac{\mathrm{d}^2\bl}{(2\pi)^2}C^{TT}_{\ell} \right]^{-1},
\end{align}

For the small-scale regime, these approximations will not significantly lose information compared to the full quadratic estimator or Bayesian estimators \cite{Bianchini_2023}.

The key challenge on small scales is separating out this signal from other non-Gaussian sky components such as gravitational lensing, the tSZ and kSZ effects and the cosmic infrared background (CIB). 
These signals all produce quadratic couplings that will contribute to this estimator and can bias measurements of $\delta \tau$. 
Analogously to lensing reconstruction, these can be removed by bias-hardening the estimator \cite{Namikawa_2013,Namikawa_2021} or by using component separation techniques to remove signals with different spectral signatures (e.g. the tSZ and CIB)\cite{Kusiak_2023}. 
These approaches will suppress these biases at the cost of reducing the signal-to-noise ratio. 
The difficulty for anisotropic screening is that the biases are much larger than the signal and thus need to be suppressed to a very high degree. 
Fully removing similar biases to lensing estimators is very challenging \cite{Madhavacheril_2018,Schaan_2019,Sailer_2021,Darwish_2021,MacCrann_2023}, and it has yet to be demonstrated that such bias mitigation methods are sufficient for anisotropic screening analyses. 
This motivates the consideration of alternative estimators that are naturally robust to foreground biases.

By examining the structure of the small-scale anisotropic screening effect \begin{align}
 T(\mathbf{\hat{n}}) \approx -T_\mathrm{large-scale}(\mathbf{\hat{n}})\delta \tau(\mathbf{\hat{n}}),
\end{align}
an estimator for anisotropic screening can be obtained by replacing the $ T_\mathrm{large-scale}$ term in \cref{eq:smallScaleQuadEstimator} with just the sign of the large scale mode and changing the normalization of the estimator. 
This gives the following estimator
\begin{align}\label{eq:newEstimator}
    \delta\tilde{\tau}(\mathbf{L}) \approx &-\int \mathrm{d}^2\mathbf{\hat{n}} e^{i\mathbf{\hat{n}}\cdot\mathbf{L}}\,\mathrm{Sign}\left[ T_\mathrm{large-scales}(\mathbf{\hat{n}}) \right]\nonumber
 T_\mathrm{small-scales}(\mathbf{\hat{n}})  \\ &\times\left[\langle| T_\mathrm{large-scales}| \rangle \right]^{-1},
\end{align}
where $\langle| T_\mathrm{large-scales}| \rangle$ is the expectation of the modulus of the large-scale field. 
It is straightforward to show that this estimator provides an unbiased estimator of the filtered $\tau$ field; see \cref{sec:implementation} for our specific filter. 
Modes with wavelengths significantly larger than those in the small-scale field will not be recovered. 
Further, for small-scale reconstructions the variance of this new estimator is only a factor of $\langle  T_\mathrm{large-scales}^2\rangle/\langle| T_\mathrm{large-scales}| \rangle^2= \pi/2$ larger. 
This result is obtained by evaluating the two expectations using a Gaussian distribution and using the relation 
\begin{align}
    \langle |T|\rangle = \int\limits_{-\infty}^{\infty}\mathrm{d}T\, | T| p(T)  = 2 \int\limits_{0}^{\infty}\mathrm{d}T\, T p(T),
\end{align}
where $p(T)$ is the probability distribution of the temperature anisotropies. 
At the cost of this extra noise, this estimator suppresses all foreground biases. 
To understand why, note that the large scales of the CMB are dominated by the primary CMB anisotropies. 
By taking the sign of this fluctuation we isolate the primary CMB from the other contributions to the large scale CMB. 
Given that the sign of $T_\mathrm{large}$ is largely clean from foregrounds, it acts as a weight with zero-average, applied to the small-scale temperature leg. 
As a result, this weighting suppresses any foreground whose sign does not correlate with the sign of the large-scale CMB.
This foreground cleaning approach is effective for the tSZ, CIB and kSZ.
A subtle bias could arise from correlations with the integrated Sachs-Wolfe (ISW) effect, which give a non-trivial contribution to the large scale modes. These can be removed by also removing the largest scale CMB modes ($\ell\lesssim 20$) or subtracting a map of the ISW field from the large scale leg. 
The latter approach is discussed in \cref{sec:robustness}.

There are similarities between our estimator and the ``gradient inversion'' lensing estimators e.g. \cite{Horowitz_2019,Hadzhiyska_2019}. 
However the gradient-inversion estimator generally increases the SNR of lensing reconstruction when lensing is dominant. 
{Screening is not the dominant source of temperature power on any scale, so } this regime does not exist for the anisotropic screening estimator, and we do not consider further dividing out the specific large scale mode, which would lead to the equivalent of the ``gradient inversion'' estimator for anisotropic screening. 
Rather our estimator can be thought of as an analogy to the shear-only lensing estimators \cite{Schaan_2019,Qu_2023}. 
The symmetry we use here to distinguish the signal from foregrounds is the changing sign of the screening effect with the large-scale CMB.
For the lensing shear estimator, it is instead the monopolar versus quadrupolar angular dependencies of foregrounds and lensing.
In both cases, we consider only part of the signal at the cost of signal-to-noise but gain robustness to foreground emission. 
A detailed comparison of these different estimators is available in Ref. \cite{Schutt_2023}.

\subsection{Lensing bias}\label{sec:lensingBias}

Lensing produces temperature anisotropies correlated with the local degree-scale CMB dipole.
Care must be taken to avoid potential lensing biases in our measurement. The importance of this effect on anisotropic screening was identified in Refs. \citep{Dvorkin_2009,Su_2011} and studied in Refs. \citep{Su_2011,Namikawa_2021,Bianchini_2023}. Refs. \citep{Sailer_2025,Hadzhiyska_2025} studied these biases for stacked estimators, as are used in this work, both analytically and in simulations.
In this section, we motivate a filtering designed to avoid any lensing bias, which we then validate in \cref{sec:validation}. 
For more discussion on the bias and filtering method, see also Ref. \cite{Schutt_2023}.
We are interested in computing the real-space correlation between $\delta\tau$ and the galaxy density field $\delta_g$, 
$\langle \delta\tau(\mathbf{\hat{n}}+\mathbf{\hat{n}}_g)\delta_g(\mathbf{\hat{n}}_g)\rangle $.
However, for simplicity of the derivation, we focus on the cross power spectrum. 
The two are simply related as
\begin{align}
   \int \mathrm{d}^2\mathbf{n}_g\langle \delta\tau(\mathbf{\hat{n}}+\mathbf{\hat{n}}_g)\delta_g(\mathbf{\hat{n}}_g)\rangle =\int \frac{\mathrm{d}^2\Bell}{(2\pi)^2} e^{i\mathbf{n}\cdot\Bell} \langle\delta\tau(\Bell)\delta^*_g(\Bell)\rangle,
\end{align}
where the integral over galaxy positions $\mathbf{n}_g$ arises as we consider the sum over all galaxy locations. 
Furthermore, we analyse the case of the estimator in \cref{eq:newEstimator} without the sign operation as this allows for a simpler analytic calculation. Note that we only omit the sign estimator when discussing lensing biases; the estimator with the sign operation is used throughout the rest of the paper.
This derivation is similar to the derivation of the lensing bias to the projected kSZ estimator \citep{Ferraro_2016} and is also presented pedagogically in \citep{Schutt_2023}. Ref.~\citep{Sailer_2025} extends this derivation to the signed estimator and finds a consistent conclusion.
We further demonstrate that this result transfers to the signed estimator using simulations in \cref{sec:validation}.
 
Under these two assumptions, we have
\begin{align}\label{eq:qe-simplified}
    \delta\tilde{\tau}(\mathbf{L})\propto&\int {\mathrm{d}^2n}{} e^{i\mathbf{\hat{n}}\cdot\mathbf{L}} T_\mathrm{small-scale}(\mathbf{\hat{n}})T_\mathrm{large-scale}(\mathbf{\hat{n}})\nonumber \\  \propto&\int_{\bl_a} 
    f_S(|\Bell_a|)f_L(|\mathbf{L}-\Bell_a|) T(\Bell_a) T(\mathbf{L}-\Bell_a)
\end{align}
where $f_S$ and $f_L$ are the high and low pass filters.

The impact of lensing is
\begin{align}\label{eq:impactOfLensing}
    T(\mathbf{\hat{n}})&=\tilde{T}(\mathbf{\hat{n}}+\nabla\phi(\mathbf{\hat{n}}))
    \approx \tilde{T}(\mathbf{\hat{n}})+\nabla\tilde{T}(\mathbf{\hat{n}})\cdot\nabla\phi(\mathbf{\hat{n}})+...
\end{align}

Thus the first-order bias from lensing is given by\footnote{We use the usual notation $\langle \rangle =\langle \rangle' (2\pi)^2\delta^{(2)}(\mathbf{L}-\mathbf{L'})$}
\begin{align}
   & \langle \delta\tilde{\tau}(\mathbf{L})\delta_g^*(\mathbf{L}')\rangle'_\text{lensing bias}=\nonumber \\&
    -\int_{\bl_a} 
    \left[f_S(|\Bell_a|)f_L(|\Bell_a-\mathbf{L}|) \mathbf{L
}\cdot(\Bell_a-\mathbf{L}) C^{\phi g}_{L'}C^{TT}_{|\Bell_a-\mathbf{L}|} \nonumber \right. \\ &\left.-f_S(|\Bell_a|)f_L(|\Bell_a-\mathbf{L}|) \mathbf{L
}\cdot\Bell_a C^{\phi g}_{L'}C^{TT}_{|\Bell_a|} \right], 
\end{align}
where the first term is from lensing contamination in the small-scale temperature leg, and the second term from contamination in the large-scale temperature leg. This result is obtained by simply inserting the lensing expansion, Eq. \ref{eq:impactOfLensing}, into the $\tau$ estimator, Eq. \ref{eq:qe-simplified}, and keeping the terms linear in the lensing potential.
In the absence of filtering or beam (i.e. $f_S = f_L = 1$), these biases vanish. 
To see this, change variables to $\ell_b=\ell_a-\mathbf{L}$ in the first term and remove the filters
\begin{align}
   & \langle \delta\tilde{\tau}(\mathbf{L})\delta_g^*(\mathbf{L}')\rangle'_\text{lensing bias}\nonumber \\=&
    -\int_\mathrm{\bl_b}  \mathbf{L
}\cdot\Bell_b C^{\phi g}_{L'}C^{TT}_{|\Bell_b|} +\int_\mathrm{\bl_a}  \mathbf{L
}\cdot\Bell_a C^{\phi g}_{L'}C^{TT}_{|\Bell_a|}\\=&
    -\int_\mathrm{\bl_b}  {L
}\ell_b C^{\phi g}_{L'}C^{TT}_{\ell_b}\cos\theta+\int_\mathrm{\bl_a}  {L
}\ell_a C^{\phi g}_{L'}C^{TT}_{\ell_a}\cos\theta^\prime.
\end{align}
The integrals over the cosines will vanish, i.e. $\mathbf{L}\cdot \Bell$ is an odd function and the rest of the integrand is even.

Now consider the case with filtering. We argue that the first term is dominant as $C^{TT}_{|\Bell_b|}\gg C^{TT}_{|\Bell_a|}$, as the CMB temperature power spectrum falls rapidly with $\ell$ and the filtering enforces $|\Bell_b|$ to be small and $|\Bell_a|$ to be large. 
The bias in the dominant first term arises from the small-scale filtering, $f_s$. Specifically, it arises as the filtering gives different weights depending on the alignment of $\mathbf{L}$ and $\bl_b$ and thus prevents the cancellation of the bias that would occur from the odd $\mathbf{L}\cdot\bl_b$.  We thus focus on the first term above.
If the filter $f_S$ were constant, this term would vanish since the integrand would be an odd function of $\Bell_b$.
For small-scale $\tau$ anisotropies, i.e., large $|\mathbf{L}|$, it is possible to pick filters such that this is true. Specifically, given a large-scale filter that is zero beyond $\ell_\mathrm{s}^\mathrm{max}$, then a small-scale filter that is constant for $\ell>L-\ell_\mathrm{s}^\mathrm{max}$ will suppress the lensing bias.

For example, if we consider $|L|>1600$ and a low-pass filter such that 
\begin{align}
   f_L(\ell)= \begin{cases}
        1 & \text{if } \ell<600, \\
        \cos\left(\pi\frac{(\ell-600)}{100}\right)  & \text{if }  600\leq \ell<650, \\
         0 & \text{if } \ell\geq 650,
    \end{cases}
\end{align}
 provided that $f_s(\ell)$ is constant for $\ell\geq950$, this first term will vanish. The size of the second term will depend on the minimum scale allowed by $f_s(\ell)$, as the steep drop off in $C^{TT}_\ell$ means that the smallest $\ell$ modes contribute most.

Since the signal part of the estimator is given by
\begin{align}
    \langle \delta\tilde{\tau}(\mathbf{L})\delta_g^*(\mathbf{L}')\rangle'_\text{signal}= &
    -\int_\mathrm{\bl_a}
    f_S(|\Bell_b-\mathbf{L}|)f_L(|\Bell_b|) C^{\tau g}_{L'}C^{TT}_{|\Bell_b|} \nonumber \\ &+\int_\mathrm{\bl_a}
    f_S(|\Bell_a|)f_L(|\Bell_a-\mathbf{L}|)  C^{\tau g}_{L'}C^{TT}_{|\Bell_a|},
\end{align}
the filtering operations will reduce the signal as well. 
However, most of the signal will be captured as the terms are weighted by $C^{TT}_\ell$, which again is largest for the small $\ell$ modes included in the low-pass filter.

In summary, if we high-pass filter the reconstructed $\tau$ map and choose the low-pass and high-pass filter widths carefully, then the leading-order lensing bias can be removed, at some cost in signal-to-noise. 
The remaining bias is expected to be suppressed relative to the signal by the ratio of the CMB temperature power spectrum in the low- and high-pass filtered legs.
In practice, we test this method on simulations of the lensed CMB to ensure that any residual lensing bias is negligible. 
An alternative approach would be to use lens-hardened estimators, as proposed in \citep{Sailer_2025}. 

\subsection{Implementation of the cross-correlation estimator}\label{sec:implementation}

To implement the estimator in \cref{eq:newEstimator}, we first need to create two filtered maps: one of the large-scale CMB anisotropies, and one of the small-scale CMB. For the low-pass maps we apply the following harmonic space filter,
\begin{align}
   f_L(\ell)= \begin{cases}
        1 & \text{if } \ell<600, \\
        \cos\left(\pi\frac{(\ell-600)}{100}\right) & \text{if }  600\leq \ell<650 , \\
        0 & \text{if } \ell\geq 650.
    \end{cases}
\end{align}
For the high-pass filter we use the following filter
\begin{align}
    f_{S}(\ell)=\begin{cases}
			0 & \text{if $\ell<850$},\\
            \sin\left(\frac{(\ell-850)\pi}{100}\right) & \text{if } 850\leq \ell < 900, \\
            1 &  \text{if $\ell \geq 900$}.
		 \end{cases}
\label{eq:high_pass_filter}
\end{align}
The structure of the high- and low-pass filters was chosen to satisfy the following criteria. First, the low-pass map needs to contain enough modes to preserve the large-scale sign. The fact that the CMB power spectrum is red means that this is satisfied provided the transition is above $\ell\sim 500$. The location of the high-pass filter was chosen such that the criteria described in the previous section are satisfied. In the squeezed limit considered here the inverse-variance filter is close to unity, and thus to simplify the estimator, we remove the inverse variance  operation. 
The smooth tapering of the filters prevents excess ringing from the harmonic space filtering. A separation between the two filters is used to prevent correlations between the high- and low-pass filtered maps. 
It is left to future work to optimize location of the filters and widths of the transitions, which could be further optimized to maximize signal to noise. Note that we do deconvolve the beam from the temperature map and deconvolve the pixel window function before applying these filters.

Next we apply a filter that removes all modes in the $\tau$ map with $\ell<1600$. 
 Together, these filters mitigate the dominant term in the first-order lensing bias, as explained in \cref{sec:lensingBias}. These filters were chosen as a balance between retaining the large-scale modes (retaining more of these modes requires filtering the $\tau$ map to higher ell) and retaining the optical depth signal (i.e. less aggressively filtering the output $\tau$ map). 
Further, to avoid the $\tau$ map being noise dominated we reapply the beam to the output $\tau$ map.

After the cuts described below, we extract a 20$^\prime$ x 20$^\prime$ cutout from the small scale map at the location of each \textit{unWISE} galaxy. To account for sky curvature we use the tangent plane projection and interpolation methods described in Ref. \cite{Schaan_2021}.
Specifically, we extract small cutouts in CEA projection (Carr\'ee Equal Area) using bilinear interpolation.
The bilinear interpolation (instead of spline) and the CEA projection help conserve flux within pixels, as verified in \cite{Schaan_2021}.
We stack these cutouts, weighted by the sign of the large scale mode, as
\begin{align}\label{eq:stackedEst}
&    \widehat{\delta\tau}^\mathrm{stacked}(\mathbf{\hat{n}}) = \nonumber \\& \frac{-\sum \mathrm{Sign}[T^\mathrm{large-scale}(\mathbf{\hat{n}}_g +\mathbf{\hat{n}})] T^\mathrm{small-scale}(\mathbf{\hat{n}}_g +\mathbf{\hat{n}})}{N_\mathrm{objects}\langle |T^\mathrm{large-scale}|\rangle },
\end{align}
where the sum is over objects locations $\mathbf{\hat{n}}_g$
and $N_\mathrm{objects}$ is the number of objects in the stack. 
We stack on 34 and 19 million  galaxies for the blue and green samples respectively; 
see \cref{tab:unWISE_prop} for source number densities. 
A 1D azimuthally-averaged profile is computed from the stacks. 
These estimators depend upon the cosmology of the primary CMB through the normalization. 
However, as the large-scale CMB anisotropies are measured at a much higher precision than the screening anisotropies, the uncertainty arising from the assumed cosmology is always subdominant. In practice, we estimate this normalization from the data itself, thereby removing this dependency.
 
The non-linearity introduced by the sign operation suppresses foreground biases in the large-scale map as the sign is determined by the primary CMB anisotropies. The immunity to foregrounds provided by the sign operation means we can include small-scale temperature measurements, which are dominated by foregrounds, without becoming biased.

We do not stack on all the objects in the input catalog. 
Instead, we perform a series of cuts to ensure that our results are robust. 
Whilst the foregrounds will average to zero over many objects, due to the sign weighting, for a finite number there can be a non-trivial residual \cite{Schaan_2021}. 
To minimize the impact of especially bright objects shifting the average, we exclude regions from our maps where there are known point sources or clusters. 
Specifically, we avoid stacking on objects that are within 6$^\prime$ of a detected SZ cluster (those in the updated Ref. \cite{Hilton_2021} cluster catalog), 6$^\prime$ of a subtracted point source (those detected via a matched filter in any of the input maps with SNR$>5$ ), 10$^\prime$ of the edge of the map  and 12$^\prime$ of sources that are inpainted. The latter are a set of especially bright objects that are inpainted during the construction of the NILC maps and are described in Ref. \cite{Qu_2022}.

Next, we ensure that in the stack there are exactly equal numbers of positively and negatively weighted regions. 
This helps further suppress the foreground emission and is implemented by selecting a random subsample -- a completely analogous approached is used in kSZ analyses \citep{Hadzhiyska_2024, Guachalla_2025}.
A heuristic of this can be understood by splitting the sky into two regions: areas where $\mathrm{Sign}[T^\mathrm{large-scale}(\mathbf{n_o})] $ is positive and areas where it is negative. Within each region foregrounds will not average down, and so contribute a bias given by the size of the average source present in the map, e.g., the average amplitude of a dusty galaxy or thermal SZ cluster. The average is obtained from many objects ($\sim 10^6$) and so will be similar in the positive and negative regions of the sky.  Thus, the degree to which these foregrounds cancel in the final estimate is set by the difference in number of sources in the positive and negative regions of the sky. The large-scale CMB has a large coherence scale ($\sim 1^{\circ}$) so in any single sky realization there will be $\mathcal{O}(10^4)$ independent positive and negative regions. The Poissionian-like scatter means we expect $1/\sqrt{10^4}\sim$1 $\%$ more sources in one region than the other. Without explicitly balancing the number of sources in each region, the degree of cancellation will leave a $\sim$1 $\%$ residual of the mean foreground signal. 
If we instead manually force the same number of sources in each region, the cancellation will only depend on the difference in the average residual in each region. This will be approximately $1/\sqrt{N_\mathrm{objects}}$  and in our case this is expected to reduce the residual by more than an order of magnitude. 

\section{Validation with Simulations}\label{sec:validation}

To validate this approach we apply the method to non-Gaussian CMB sky simulations. The goal is to verify both that the other sky signals do not bias the estimator and that the estimator is able to recover the input signal. These tests are structured to approximate the sky as observed by ACT and the \textit{Planck} satellite so that they serve as a validation of the data analysis presented in the next section. 

\subsection{Simulation Properties}
We primarily consider two non-Gaussian sky simulations: the Websky \cite{Stein_2020} and Agora \cite{Omori_2024} simulations. Each simulation contains the following  correlated and non-Gaussian CMB secondary anisotropies: the tSZ, kSZ, CIB, lensing and radio galaxies. Both simulations are dark matter only simulations with the secondary anisotropies then ``painted'' on in postprocessing. As the two simulations use distinct methods and models for this painting procedure, they provide some measure of the theoretical uncertainty on the properties of the CMB secondaries. Additionally, we use the \textsc{backlight} simulations, an upcoming new suite of non-Gaussain sky simulations, to analyze potential biases from gravitational lensing. From this suite of simulations, we only use the CMB lensing and dark matter halo catalogs. This suite of simulations has many sky realizations (the full suite will have >1000 simulations, but we used the 40 simulations available at this time), unlike the others which only have one, making it ideal for testing for biases. 

Using these simulations of the sky, we construct simulated observations that match the properties of the ACT and \textit{Planck} data sets used in this work. These simulations match the instrumental properties including map noise, beams and observation frequencies. We refer the reader to Ref. \cite{Atkins_2023} and Appendix B of Ref. \cite{Coulton_2023} for a thorough description of these simulations. We use the Needlet Internal Linear Combination (NILC) component separation pipeline, described in Ref. \cite{Coulton_2023}, to combine simulations of the individual observations into a map of the CMB anisotropies. Through this procedure we obtain maps with the appropriate levels of residual foregrounds that are present in the observations.

The Websky and Agora simulations do not include any anisotropic screening signal. To validate that our method provides an unbiased measurement of non-zero signals, we created an additional version of the simulations that contains an approximate anisotropic screening signal. We first construct an approximate $\tau$ map, by rescaling the Compton-$y$ map from each simulation by a constant factor to ``convert'' from Compton-$y$ to optical depth. This factor is an approximation of the mean gas temperature of the halos and is chosen to give an average $\tau$ similar to the expected value ($\sim 5 \times 10^{-4}$). Note that this is a crude approximation, and thus is not a prediction for the optical depth in each simulation.
With this approximate $\tau$ map we create the anisotropic screening effect by using it to screen the primary CMB anisotropies, via \cref{eq:screeningEffect}.

The catalogs used in the stacking analysis are the halo catalogs of the two simulations. For the Websky, \textsc{backlight} and Agora et al. simulations we used a halo mass cut that ensures that we have a similar number of halos to galaxies used in the data analysis ($M_h^\mathrm{min} \approx 9 \times 10^{12}$).
As in the processing of the data, we avoid stacking on objects that are close to point sources, clusters or the edge of the mask. For the data, the sources are identified in the maps using point source and cluster finders, e.g.,\cite{Datta_2016,Gralla_2020,Hilton_2018,Hilton_2021}; however, for the simulations, we instead use the true source catalogs. The cut thresholds are set to approximately match the point source and cluster thresholds of the ACT data: CIB sources with amplitudes $\gtrsim 350 \mu$\,K ($>30$\,mJy) at $217\,$GHz, radio sources with amplitudes $\gtrsim 150 \mu$\,K ($>9$\,mJy) at $90\,$ GHz and clusters with $M_{200}\gtrsim 5\times 10^{14}$ M$_\odot$/h.

\subsection{Assessing foreground biases with Agora and Websky}

\begin{figure*}
    \centering
  \subfloat[Websky simulations] {\label{fig:websky_sims} \includegraphics[width=0.48\textwidth]{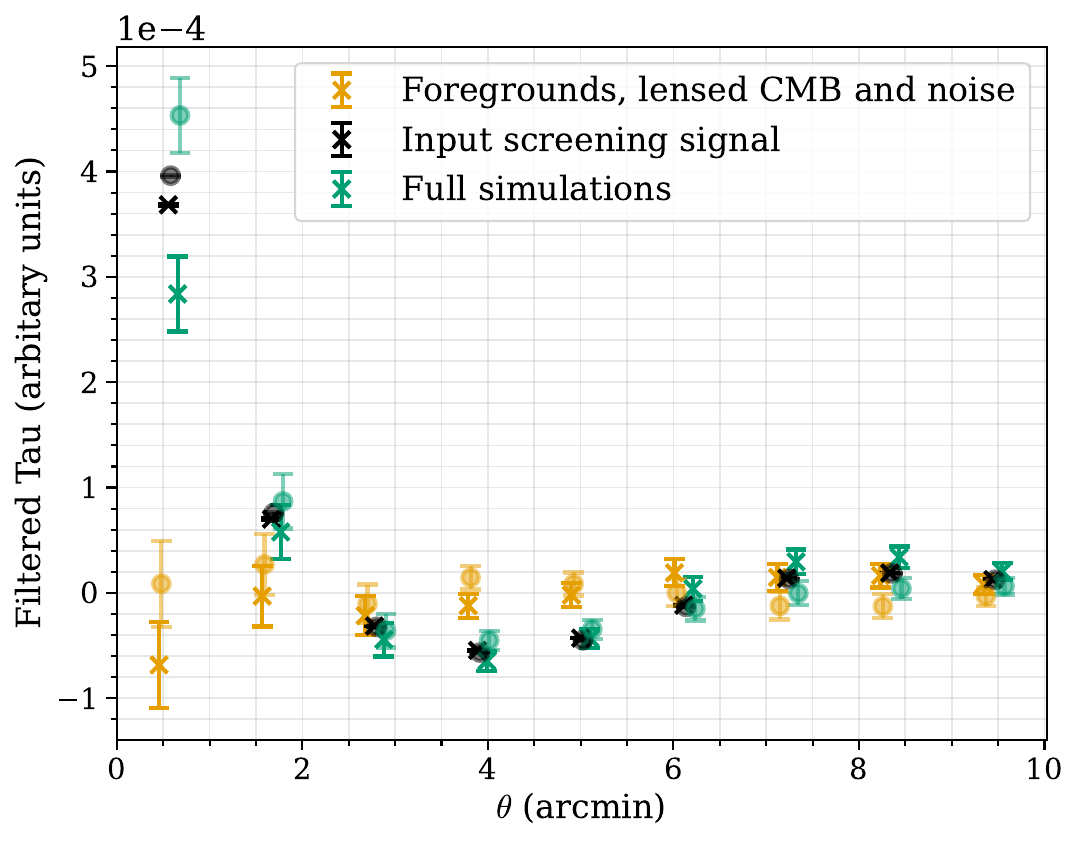} }
   \subfloat[Agora simulations] {\label{fig:agora_sims} \includegraphics[width=0.48\textwidth]{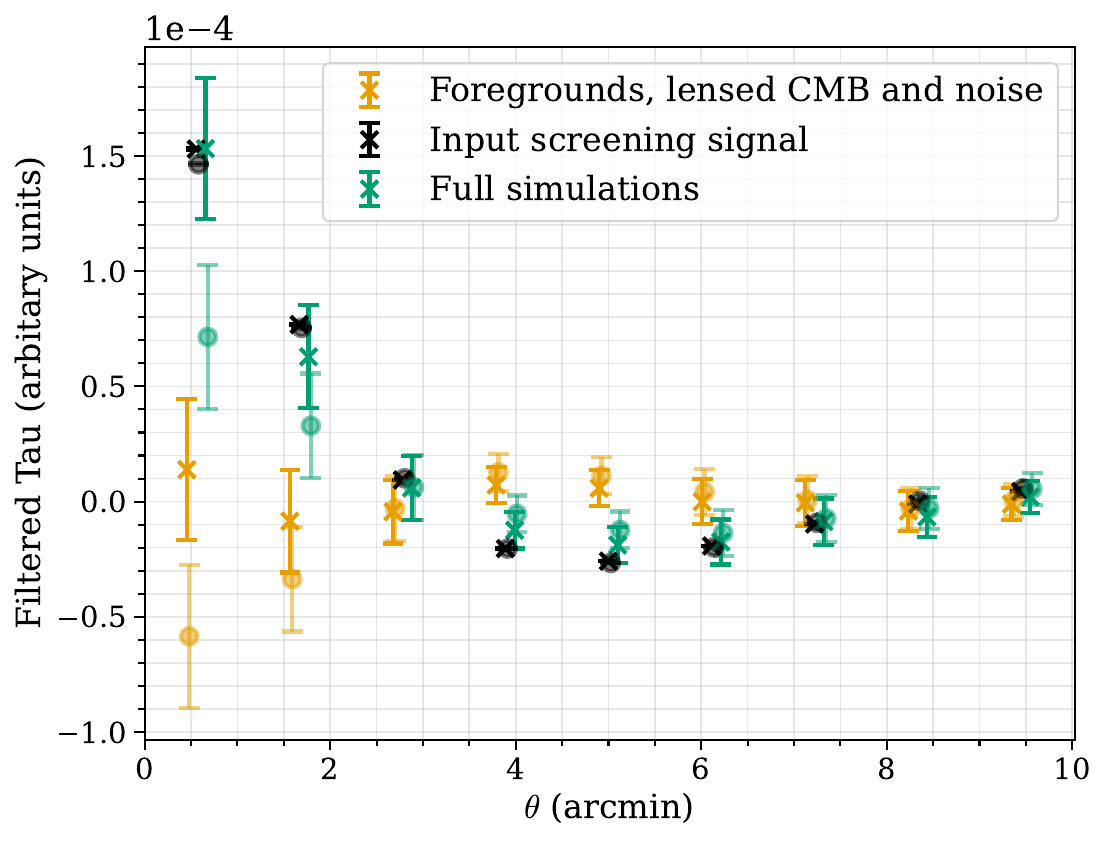}}
 \caption{ 
 An application of our stacked estimator to mock ACT \& \textit{Planck} observations constructed from the Websky (left) and Agora (right) simulations. 
 The cross and circle points correspond to the two realizations. 
 We apply the estimator to simulations that contain foregrounds but no anisotropic screening signal (orange lines). These are consistent with zero (see \cref{tab:results_chisqu_null}) and demonstrate that our method is immune to foreground biases. We also show that the estimator is unbiased, as the measurements on simulations with a $\tau$ signal (green) match the input signal  when filtered in the same manner (black).  All error bars denote $1\sigma$.}
\end{figure*}
In \cref{fig:websky_sims} and \cref{fig:agora_sims} we apply the method to the Websky and Agora simulations,
matching the ACT maps and unWISE catalog as closely as possible. As the ACT footprint covers less than 1/2 the sky we can extract two approximately independent realizations of the sky from each of the simulations. The input signal is measured by running the estimator on two maps: a large scale map containing only the CMB and a small scale map containing only the screening anisotropies. The other analyses include all the potential contaminants: the tSZ, kSZ, CIB, radio galaxies and CMB lensing.
As can be seen when applied to the default simulations with foregrounds but without any anisotropic screening signal, the estimator measurements are consistent with zero. 
This demonstrates that our method is not biased by the foregrounds present in these simulations. 
When applied to the simulations with a mock screening signal, we recover measurements that are consistent with the input $\tau$ signal. The oscillatory structure seen in the signal arises from the high-pass filtering performed on the small-scale CMB map.

These two tests provide evidence that our pipeline is able to provide robust and unbiased measurements of a anisotropic screening signal.

\begin{figure}
\centering\includegraphics[width=0.48\textwidth]{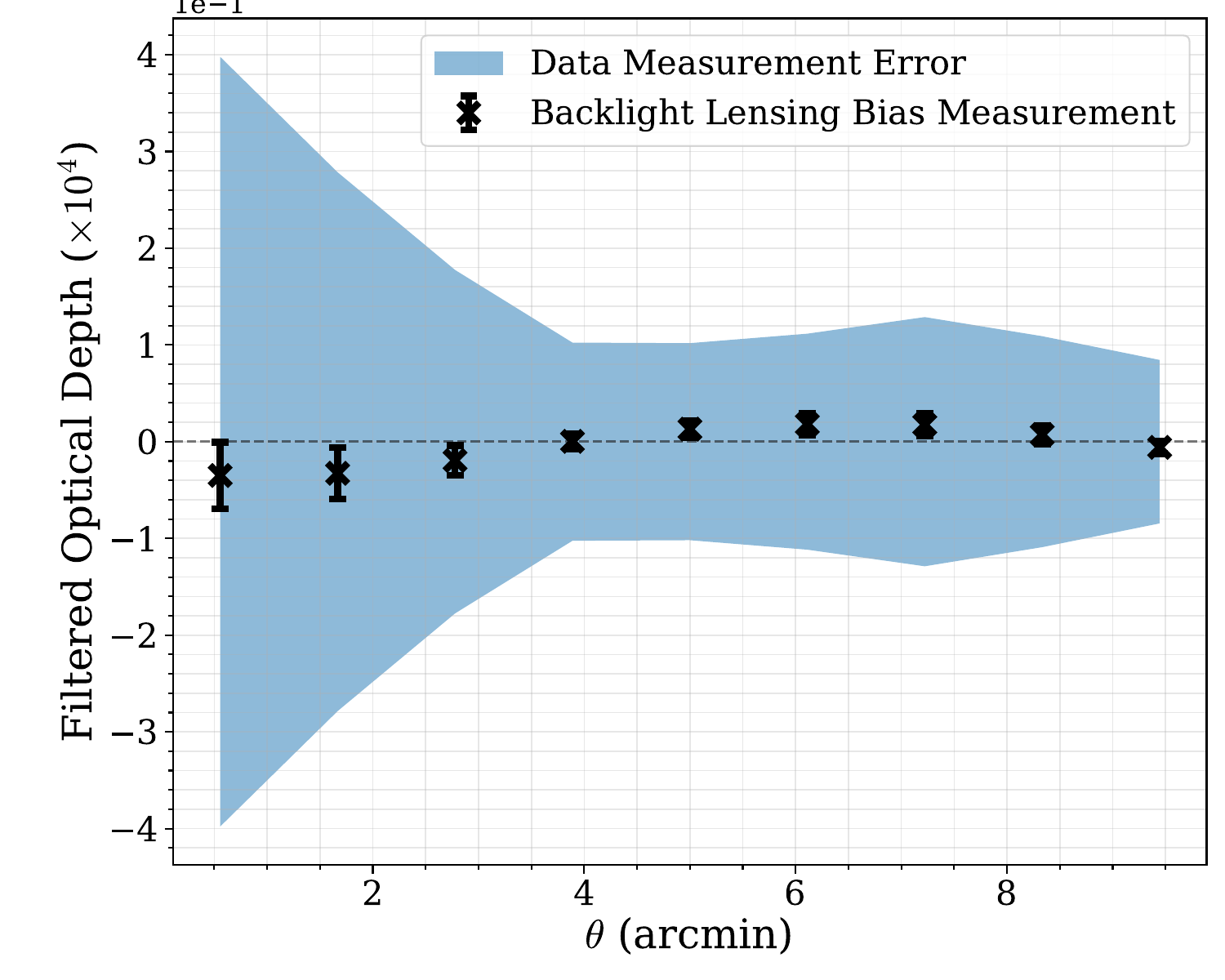} 
 \caption{ The bias of CMB lensing on our analysis method as assessed by performing our analysis on lensed CMB maps from the \textsc{backlight} simulations. 
 The resulting mean optical depth profile, shown in black, is consistent with zero, given the 1$\sigma$ error on the mean. 
 For comparison we shown the data measurement error bars, centered around zero; these errors are much larger. Together these results imply that CMB lensing does not bias our estimator. 
\label{fig:backlight}}
\end{figure}

\subsection{Assessing the lensing bias with the \textsc{backlight} simulations}

As discussed in \cref{sec:lensingBias}, CMB lensing can potentially lead to a large bias in the measurement. 
In this section, we validate that our filtering methodology is effective in mitigating the lensing bias. We use the \textsc{backlight} simulations for this purpose as there are many realizations of the sky and this allows us to reduce the measurement error and to better differentiate a noise fluctuation from a bias. 

For each of the 40 simulations we perform our analysis on a lensed CMB map at the location of dark matter halos. 
In \cref{fig:backlight}, we show the resulting mean optical depth profile. 
It is consistent with zero and much smaller than the statistical uncertainty on the actual measurement, showing that our estimator is not biased by CMB lensing. 
This result is non-trivial as our derivation showed there was a second term, which we argued is subdominant, and the derivation is without the complications of the sign operation. This simulation based validation increases our confidence. 

This measurement also constrains the size of a mean-field term. Mean field terms arise in quadratic estimators from statistical anisotropy, primarily from the mask (see e.g.,\citep{Namikawa_2014} for further details). Our measurement is consistent with zero and thus the mean field is small. A similar result obtained with our noisy simulations, used for the covariance matrix described in \cref{sec:uncertainity}, further supports these conclusions.

\section{Data Sets}
\label{sec:data}

The anisotropic screening anisotropies, \cref{eq:screeningEffect}, are arcminute-scale fluctuations that trace the small-scale optical-depth fluctuations. 
Detecting the signal thus requires measurements of the small-scale CMB anisotropies. 
The high resolution and sensitivity of current and future ground-based CMB experiments make them ideal for this task. 
In this work we combine CMB data from the ACT Data Release 6 and the \textit{Planck} satellite \cite{planck2016-l01} to extract the screening signal. 
We use a component separation technique to combine observations at multiple frequencies and produce a map of the CMB and screening component, as detailed in  Ref. \cite{Coulton_2023}. 
This is a first step in removing contaminant signals.
Here we briefly review the key products and refer the reader to  Ref.~\cite{Coulton_2023} for more details.

The \textit{Planck} satellite observed the full sky over four years with an angular resolution of $7.22^\prime$ at $143\,$GHz \cite{planck2016-l01}. 
The component-separated maps are constructed from the NPIPE \textit{Planck} data release \cite{Planck_int_LVII} and use data between $30\,$GHz and $545\,$GHz. 
These maps provide the large-scale CMB measurement; the small scales come from high-resolution measurements of the millimeter sky by ACT. 
ACT observed about one third of the sky over the course of 15 years \cite{dunner/etal:2013,dunkley/etal:2011,das/etal:2011,sievers/etal:2013,gralla/etal:2020,louis/etal:2017,naess/etal:2014,Aiola_2020,naess/etal:2020}. 
This work uses data from ACT Data Release 4 (DR4) and Data Release 6 (DR6), which contain data from 2013 t0 2022. 
It comprises observations at three frequencies:  98, 148, and 225 GHz (known as f090, f150, f220). 
At $148\,$GHz the ACT beam full-width-half-maximum (FWHM) is $1.4$$^\prime$. Typical noise levels in the ACT maps are 13/12/48 $\mu$K arcmin at f090/f150/f220 \citep{Naess_2025}. 
Atmospheric noise, primarily from the vibrational and rotational modes of water vapor, has a red spatial power spectrum and inhibits ACT from making observations of the large scale CMB anisotropies \cite{Morris_2022}. Thus the combination of \textit{Planck} and ACT data is well suited to studying the anisotropic screening signal.

The ACT and \textit{Planck} multifrequency observations are combined via a component separation method called the needlet internal linear combination (NILC) method, to produce a map isolating the CMB anisotropies. 
This CMB map is dominated by the primary CMB anisotropies on angular scales ranging from full-sky to a few arcminutes. 
The resulting map is convolved to a 1.6$^\prime$ FWHM Gaussian beam. 
The NILC map we use here was optimized to produce an unbiased CMB map with minimal ``noise'' variance, with noise defined to be all signals (sky or instrumental) that are not the component of interest, in this case the CMB \cite{Bennett1992,bennett2003b}. 
This method reduces foregrounds, but does not eliminate them completely; hence the need for the ``signed'' estimator and extensive foreground tests in \cref{sec:robustness}.  
The robustness of our measurement can be tested by making variants of the NILC maps. 
These variants are produced by either using only subsets of the data during the NILC process (if only one frequency is used the NILC method can be thought of as a means of coadding data) or by applying constraints to ensure that a known contaminant signal does not contribute to the map ``noise.'' To achieve this latter goal, we use the constrained NILC method \cite{Chen_2009,Remazeilles_2011b}.

\begin{table}[]
    \centering
    \begin{tabular}{c|c|c|c}
        Sample & Mean Redshift & Mean Halo Mass  & Source density  \\
        & & (M$_\odot$/h)& (arcmin$^{-2}$ )\\ \hline\hline
        Blue &0.6& $ 1.4 \times 10^{13} $ &  0.58\\ 
        Green &1.1& $ 1.3 \times 10^{13} $& 0.32 \\
    \end{tabular}
    \caption{ The key properties of the \textit{unWISE} galaxies as measured by Refs \cite{Krolewski_2020,Kusiak_2022,Kusiak_2023}. The source density is computed after our catalog cuts. }
    \label{tab:unWISE_prop}
\end{table}

CMB experiments are only sensitive to the anisotropic screening signal integrated along the line of sight. 
To understand the host mass and redshift dependence of the signal, we cross-correlate the CMB data with another tracer of the gas distribution \cite{Roy_2022}. 
If the second tracer is measured at high statistical significance, then cross-correlation-based methods also provide a means of boosting the detectability of the anisotropic screening effect. 
In this work we cross-correlate the ACT-\textit{Planck} CMB measurements with galaxies in the \textit{unWISE} catalog \cite{Krolewski_2021}.
 
The WISE satellite observed the sky from 2010 to 2011 and then again from 2013 to the present. We use the ``blue''  and ``green''  galaxy samples\footnote{We do not consider the ``red'' sample as it has a significantly lower number density and so significantly lower constraining power.} from Ref.~\cite{Schlafly_2019,Krolewski_2020}\ and refer the reader to Ref.~\cite{Krolewski_2020} for details on the selection criteria. 
These samples, whose properties are summarized in \cref{tab:unWISE_prop}, are obtained from the deep 3.4 and 4.6 micron \textit{unWISE} catalogs \cite{Meisner_2017a,Meisner_2017b,Meisner_2018}, with \textit{Gaia} data used to help remove contaminant stars \cite{Gaia_2018}. 
The high source densities and large overlap with ACT are the key reasons for using the \textit{unWISE} catalogs.
Moreover, in the case of anisotropic screening, the lack of accurate individual redshifts does not reduce the signal-to-noise.

\section{Uncertainty quantification}\label{sec:uncertainity}

\begin{figure}
\centering\includegraphics[width=0.48\textwidth]{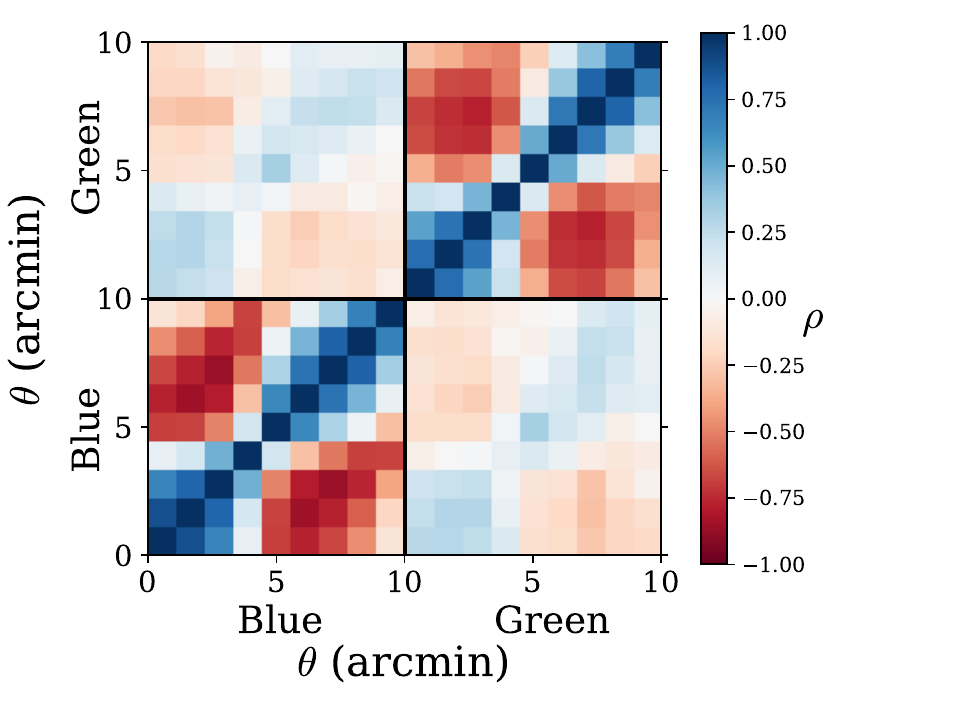} 
 \caption{ The correlation matrices of the $\tau$ measurements from the \textit{unWISE} blue and green samples, computed from 150 independent simulations that do not include the anisotropic screening signal. The lower left corner of the plot is the blue sample, the upper right is the green sample and the diagonals are the cross terms. Significant off-diagonal correlations arise from the high-pass filtering. No strong correlations are detected between the two samples, which implies that the measurement noise is uncorrelated between the samples.
\label{fig:correlationMatrix}}
\end{figure}

To compute the errors,  we generate a set of 150 simulated Gaussian temperature maps, matching the properties of our real NILC map.
As the anisotropic screening signal is highly subdominant to all the other sky signals, we can simply treat the NILC map as a map of ``noise.''
This is used as the input to the tiled noise model, described in Ref. \cite{Atkins_2023}. With this formalism we can accurately capture the variation in the ``noise'' properties in both scale and location across the map. From these we generate an ensemble of 150 simulations. 
For each simulation we perform the same filtering as the data and then apply the stacked sign estimator at the locations of the \textit{unWISE} galaxies. 
The weighting is given by the sign of the large-scale modes of the CMB map. 
This is repeated over the ensemble of simulations to obtain an estimate of the covariance matrix. 
The error bars in all plots show 1$\sigma$ errors computed via this approach. 
The $\chi^2$, $p$-values and the model comparisons quoted throughout the paper also use the full covariance matrix described here.

 In \cref{fig:correlationMatrix} we show the correlation matrix for the blue and green samples. The strong radial correlations induced by the filtering can clearly be seen. These correlations are not due to the signal -- the simulations do not include the anisotropic screening signal -- and arise from the primary CMB, noise and other sky components. These signal-free simulations quantify the expected noise-induced variance in the profiles. The comparison of our measurement with this noise quantifies the significance of the measurement. 
 There is evidence for correlations between the two samples. This correlation does not need to be modeled when quantifying the detection significance, as we are then simply assessing the rejection significance of the null hypothesis, in which there is no screening signal anyway. However, any analyses using a combined sample should account for this effect.

As a rough sanity check for this approach,
we compute an approximate covariance matrix from the data themselves by computing the covariance matrix from the scatter across the individual $\tau$ measurements from each galaxy. 
If each galaxy's profile were independent, the covariance matrix of the stack would be $1/\sqrt{N_\mathrm{objects}}$ times the single galaxy covariance matrix. 
In practice the galaxies are not  independent, as the source density is high and the ``noise'' to screening --i.e. all other signals in the map-- is not white, and thus this covariance matrix would slightly underestimate the true variance. We found that this approximate covariance matrix was qualitatively in agreement with our simulated covariance matrix ($\sim35\%$ agreement between the diagonal elements of the two matrices), providing a rough validation of our simulations.

\begin{figure}
\centering\includegraphics[width=0.50\textwidth]{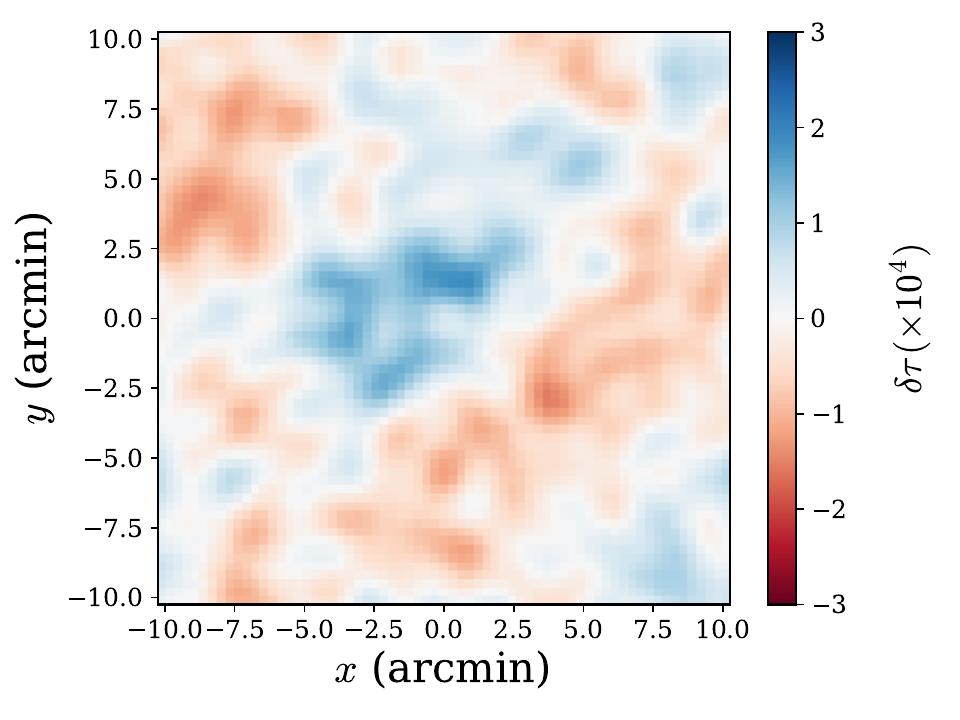}
 \caption{
 The 2D stacked optical-depth map around the \textit{unWISE} blue galaxies. This high-pass filtered optical-depth map is obtained by running the ``sign'' estimator on ACT data at the positions of the galaxies.
The optical-depth map is convolved with both a high pass filter and the ACT beam (a Gaussian with a 1.6$^\prime$ full width). The expected screening signal from the \textit{unWISE} galaxies would appear as a very bright blue blob in the center of the map.
 \label{fig:2Dstack} 
 }
\end{figure}

\section{Anisotropic Screening Constraints}
\label{sec:results}

\begin{figure}
    \centering\includegraphics[width=0.48\textwidth]{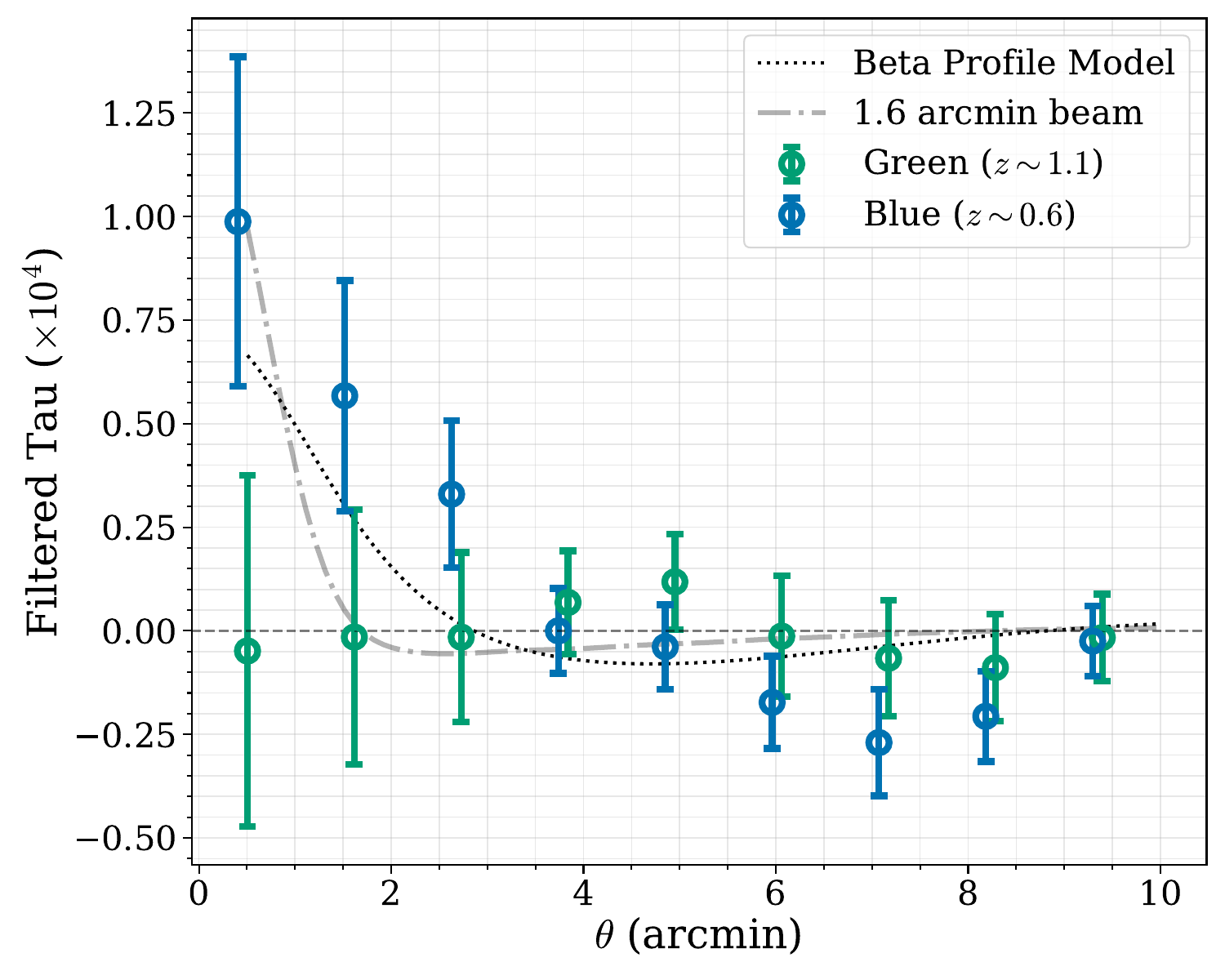} 
 \caption{Stacked 1D profiles, with $1\,\sigma$ error bars, of the optical depth around the two different unWISE samples, green and blue. 
 These are obtained by azimuthally averaging the 2D stacks.
 The dotted line is a beta profile model of the electron density, with parameters fit to the  blue sample. As context, we show the expected signal for a point source, the dot-dashed line.
 The point-source and beta profiles are convolved to the beam (1.6$^\prime$ FWHM) and filtered in the same manner as the data. 
The error bars are highly correlated across angular bins, consistent with \cref{fig:correlationMatrix}.
 \label{fig:data}}
\end{figure}

In \cref{fig:2Dstack} we present the 2D stack of the sign-weighted CMB data on the unWISE blue sample, 
where the anisotropic screening signal is not readily apparent.
We show the azimuthally averaged 1D profiles of optical depth, computed with \cref{eq:stackedEst}, for the two \textit{unWISE} samples in \cref{fig:data}. 
We see no strong evidence of a signal in either the blue and green samples.
The ``ringing'' structure in the profiles is a result of the filtering applied to the CMB map.

\subsection{Measurement significance}
\label{sec:significance}

For the blue and green samples, \cref{tab:results_chisqu} provides $\chi^2$ between the data and the null hypothesis, and the corresponding $p$-values. 
The final column translates these $p$-values to the equivalent number of Gaussian sigmas. 
These show that our measurements are consistent with the null hypothesis, i.e. with a non-detection.
Our measurements can still be used to provide a limit on the maximum optical depth associated with the two samples.

We fit the data to a two-parameter phenomenological model, a 2D symmetric beta profile \cite{Cavaliere_1978}.
Specifically, we assume the 3D electron distribution follows a beta profile as
\begin{align}
    n_e(\mathbf{r}) = n_e^0 \left(1+(r/r_0)^2\right)^{\frac{-3\beta}{2}}
\end{align}
where $n_e^0$ is a normalization, $r_0$ is the scale radius and $\beta$ characterizes the slope of the profile. Integrating along the line of sight gives the projected profile as
\begin{align}
    \tau(\mathbf{\theta})=\tau_0 \left(1+\frac{\theta}{\theta_0}\right)^{(1-\frac{3\beta}{2})},
\end{align}
where $\tau_0$ is the central amplitude and $\theta_0$ is scale radius. Due to the degeneracies between $\theta_0$ and $\beta$, we follow \cite{Plagge_2010} and fix $\beta=0.86$.
 The beta profile is then convolved with the beam and filtered in a manner identical to the data. 

\begin{figure}
    \centering\includegraphics[width=0.48\textwidth]{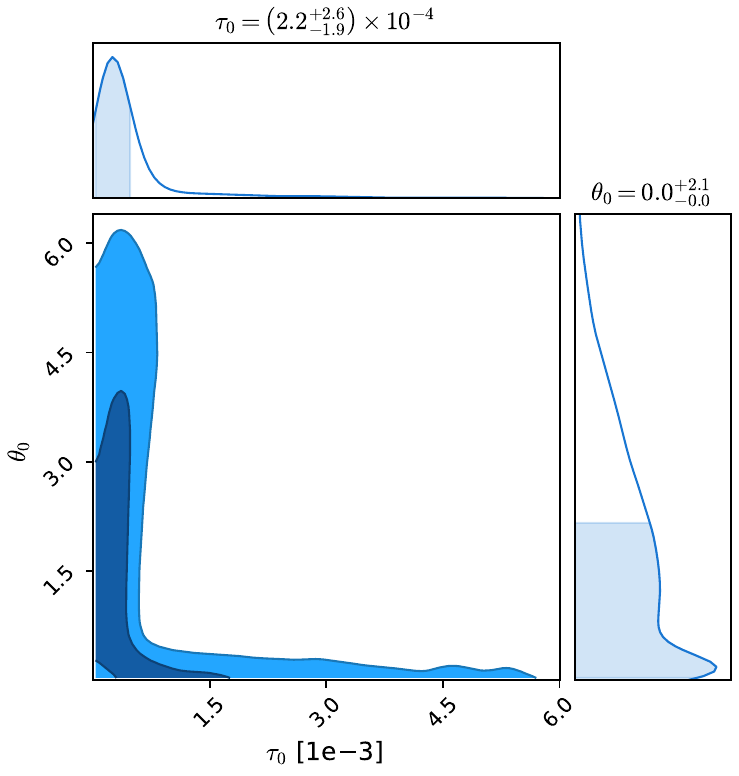} 
 \caption{The best fit beta profile parameters to the unWISE blue sample. We find that the amplitude is constrained to be $\tau_0=2.2_{-1.9}^{+2.6}\times 10^{-4}$ or $\tau_0<2.1\times10^{-3}$ at 95\%.
 \label{fig:2D_contours}}
\end{figure}

 Using a Gaussian likelihood with broad flat, positive definite priors on the amplitude and width, we use \textsc{emcee} \citep{Mackey_2013} to sample the posterior.
The best fit profile is plotted in \cref{fig:data},  and its $\chi^2$ statistics are in \cref{tab:results_chisqu}. The $\delta\chi^2$ between our best fit model and the null hypothesis is $5$, indicating no strong evidence of a detection. Note that the best fit values, i.e. minimum chi squared, are $\tau_0=1.9 \times 10^{-4}$ and $\theta_0= 1.2$ and are different from the 1D posterior means shown in \cref{fig:2D_contours}.
These two parameters provide a useful compression and summary of our measured data points, which can be used for future modeling endeavors. 
They also give us an additional estimate of the measurement significance. Finally, by integrating this profile out to 6\,arcmin we can constrain the total optical depth, $ \bar{\tau}$. We find the $\bar{\tau}< 0.033$ arcmin$^2$ at 95\% confidence.\footnote{Note that as we integrate the optical depth (dimensionless) over a disc (units arcmin$^2$) our measurement is in units of arcmin$^2$.} Repeating this analysis for the green sample leads to an upper bound on the integrated optical depth of $\bar{\tau}< 0.057$ arcmin$^2$ at 95\% confidence.

We consider two statistical tests to compare the beta profile to the null hypothesis: the Akaike Information Criterion (AIC) and an F-test \cite{Akaike1998,Akaike1974,Hahs-Vaughn2012}. 
We find AIC values of $-0.93$ and $4.0$ for the blue and green samples, respectively, corresponding to no strong evidence for the beta model.  
The F-test also yields no strong evidence for a nonzero signal, with $p=0.25$ and $1.0$, respectively.
\begin{table}
\begin{tabular}{ c |  c |  c | c  }
  \multicolumn{4}{c}{ Rejection of null hypothesis}  \\ 
 \hline\hline
 Sample & $\chi_\text{null}^2$ & $p$-value  & $N_\sigma$ \\   
\hline\hline
 Blue & 15.3 & 0.08& 1.7 \\ 
Green &2.2&0.98 & 0.016 \\
\hline\hline
\end{tabular}
\begin{tabular}{ c |  c |  c  }
\multicolumn{3}{c}{Goodness of fit for the 2 parameter beta profile }  \\ 
 \hline\hline
 Sample & $\chi_\text{best fit}^2$ & $p$-value   \\  
 \hline\hline
 Blue & 10.3 & 0.46  \\ 
Green & 2.2& 0.98  \\
\end{tabular}
\caption{
Assessment of the detection significance. The $\chi^2$, $p$-values and equivalent Gaussian sigma significance ( $N_\sigma$) for the ``no signal'' null hypothesis (left) and the best-fit beta profiles (right). 
\label{tab:results_chisqu}}
\end{table}

\subsection{Robustness Tests}\label{sec:robustness}

\begin{figure}
\centering\includegraphics[width=0.5\textwidth]{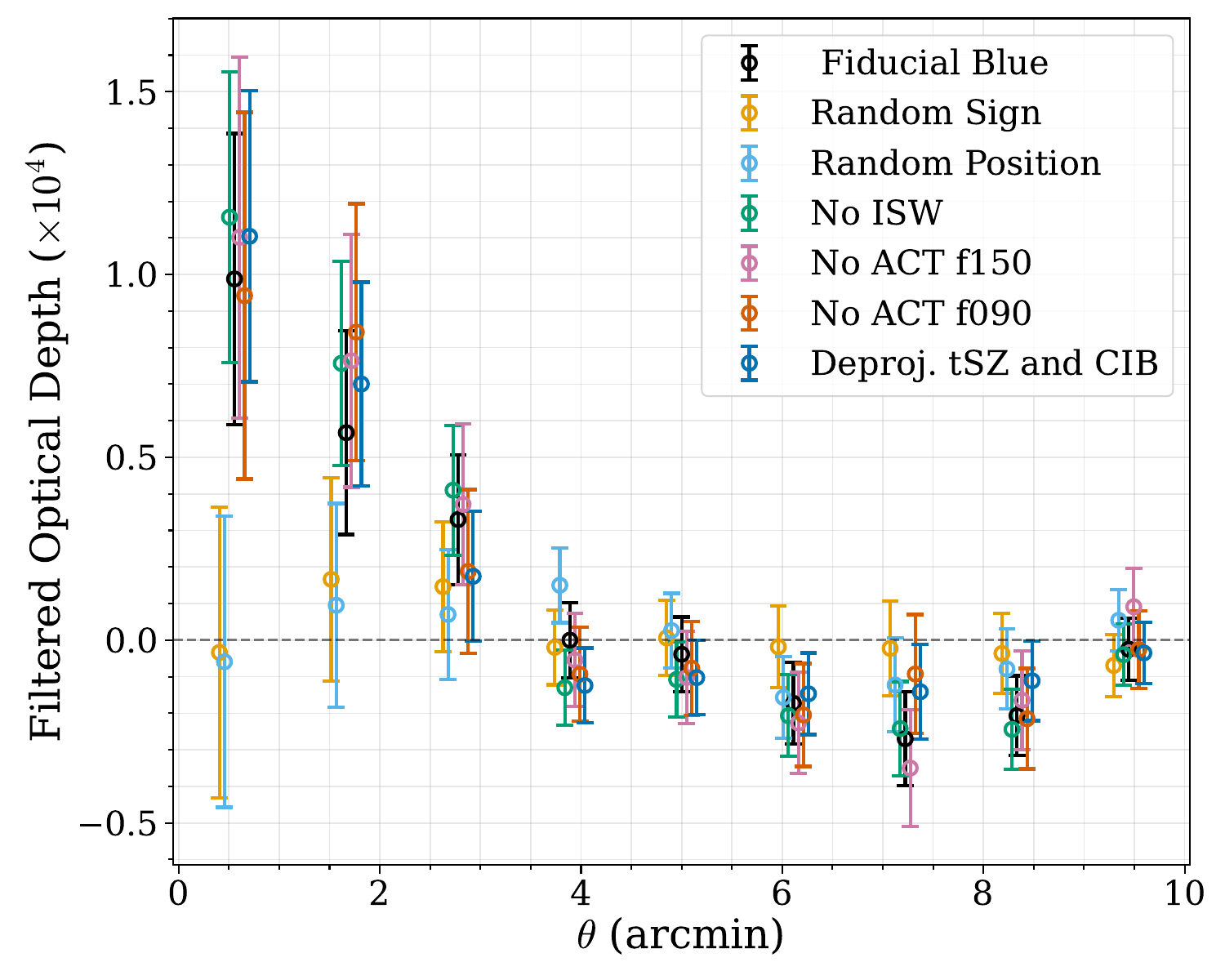} 
 \caption{A summary of the null and consistency tests performed to validate our result. The two null tests, stacking on random location and randomizing the sign used in the stack, are both consistent with zero and provide a data-based check of our method (see \cref{tab:results_chisqu_null}). The two foreground tests, which are labeled ``Deproj. tSZ and CIB'' and ``ISW removed'',  explicitly remove contamination from the tSZ, CIB and ISW effect. These demonstrate that our result likely contains no significant biases from these effects. The final tests, using only data from 93GHz or 148GHz, are consistent with anisotropies that have a blackbody spectrum, as expected for the anisotropic screening effect. All error bars denote $1\sigma$. Note that the error bars are not significantly increased by deprojection as we only deproject the tSZ and CIB in the large-scale leg.
 \label{fig:dataNulls}}
\end{figure}

\begin{table}
\begin{tabular}{ c |  c |  c | c  }
 & & &  Equivalent  \\
  & $\chi^2$ & $p$-value  & Gaussian $\sigma$ \\  
\hline\hline
Random positions  & 5.5& 0.79 & 0.26 \\ 
Random signs  & 12 &0.22 &1.2  \\
\end{tabular}
\caption{The $\chi^2$, $p$-values and equivalent Gaussian $\sigma$ for two of the null tests. Both null tests have 9 degrees of freedom. 
\label{tab:results_chisqu_null}}
\end{table}

To validate our results, we perform a series of data-based robustness and null tests.

We performed two null tests which constrain potential foreground biases and validate our analysis pipeline.
First we repeated our stacking analysis on random locations. Our estimator, \cref{eq:stackedEst}, is not sensitive to the mean optical depth and so it is expected that if we stack on random locations we should see a signal consistent with zero. The number of locations is the same as the number in the original catalog, for which we use the largest sample, the blue sample.
The measurement using random locations is also shown in \cref{fig:dataNulls} (light blue).
The $\chi^2$ of this null test compared to zero, reported in \cref{tab:results_chisqu_null}, is consistent with no signal. 
This tests our estimator, our covariance matrix and mean field assumptions.

Second, we repeat our stacking analysis, but randomize the sign used to weight each profile. 
This test should null the screening signal, but could potentially reveal foreground biases if present.
Foregrounds are expected to average to zero in our standard estimator. 
However, if there are any anomalously large values they may average down very slowly and bias our result. 
This ``random sign'' test should identify such cases as randomizing the sign will remove the anisotropic screening signal. 
As seen in \cref{fig:dataNulls} (yellow) and \cref{tab:results_chisqu_null}, this null test is also consistent with no signal.

Next we perform tests aimed at identifying contamination in our measurement from two known sources: the thermal Sunyaev-Zeldovich effect (tSZ) and the cosmic infrared background (CIB).
To test for biases from the tSZ and CIB, we make a constrained NILC map that explicitly removes the tSZ and CIB effects as described in Ref.~\cite{Chen_2009,Remazeilles_2011b}.
This map is referred to as the CIB- and tSZ-deprojected map; we refer the reader to \cite{Coulton_2023} for a detailed description of our implementation. 
This operation results in a significant increase in the noise in the component-separated CMB map.
We could deproject the tSZ and CIB separately. 
However, as shown in \cite{Sailer_2021,Coulton_2023a}, removing only one component often increases the other component, so it is hard to predict how our results would change.
The large loss of statistical power when using the deprojected map can be partially avoided, by only using this map for one of the legs of our estimator.
To remove any biases from our measurement we note that we only need to remove the contamination from one of the two maps used in our estimator. 
In this case we can use the tSZ- and CIB-deprojected map as the low-pass filtered map for the sign operation. 
We make this choice as the scales relevant for the large-scale sign are largely unaffected by the increased noise in the deprojected map because they are dominated by the primary CMB modes. 
The resulting measurement, shown in \cref{fig:dataNulls}, is consistent with our baseline measurement. Together these results increase our confidence that we do not have a large tSZ or CIB bias.

We also check for contamination from the ISW effect, which could lead to a bias in our estimator. 
This is because the ISW contributes a significant amount of power to the CMB map on large-scale, correlated with small-scale modes, and cannot be removed by multifrequency cleaning. 
The ISW could be biasing the large-scale sign.
The ISW effect traces the large scale potential field and so is correlated with other cosmic structures, such as the tSZ, CIB and \textit{unWISE} galaxies, and thus could produce a signal in our analysis. Explicitly, the bias would arise from the ISW-galaxy-small-scale foregrounds bispectrum.

It is expected that any bias from the ISW effect will be small as, even with optimal estimators, it is difficult to detect the ISW effect at high significance \cite{Stolzner_2018,Amendola_2018}. 
To test for this we subtract the \textit{Planck} collaboration's map of the ISW effect (see Ref.~\cite{planck2014-a26} for details of the construction of this map) from our component-separated CMB map. We then repeat our stacking analysis. 
By explicitly removing the ISW effect we should mitigate any potential bias.
As shown in \cref{fig:dataNulls}, our measurement is largely unchanged. 
This suggests that our measurement is not dominated by bias from the ISW effect. 
This null test has two limitations: first, it is difficult to produce a map of the ISW anisotropies. The map produced in Ref. \cite{planck2014-a26} is noisy and not a perfect tracer of the true ISW effect. 
Thus our subtraction will not perfectly and noiselessly remove any biases. However, the ISW effect is only likely to bias our measurement on the largest scales, where it may impact the measured sign, and these are the scales where the ISW effect is best measured. 
Second, the ISW map has a slightly different mask compared to our fiducial analysis (see Fig. 20. of Ref.  \cite{planck2014-a26}). 
This will introduce some noise and scatter in our measurements. Given the very small impact of subtracting the ISW map, we expect both of these effects to be minor. 
An alternative method would be to filter out the largest scales. We do not consider that approach here as the filtering would add large amounts of noise to the sign measurement.

The final test performed is aimed at characterizing the spectral response of the observed signature signal. 
The frequency dependence of the anisotropic screening signal is expected to be the same as the primary CMB anisotropies. 
In the differential thermodynamic units used in this analysis, the anisotropies are expected to be the same size at all frequencies. 
We test this by generating maps that contain only observations at $93\,$GHz or $148\,$GHz. 
These maps are then high-pass filtered and used in the estimator and converted to the same beam.
These two measurements will be consistent if the observed signal has no frequency dependence. 
We choose these two frequencies as they have sufficiently low noise and high resolution to obtain a high SNR measurement. 
\cref{fig:dataNulls} shows that these two measurements are consistent and thereby provides evidence that our observed signal has the expected frequency dependence.

\begin{figure}
    \centering\includegraphics[width=0.48\textwidth]{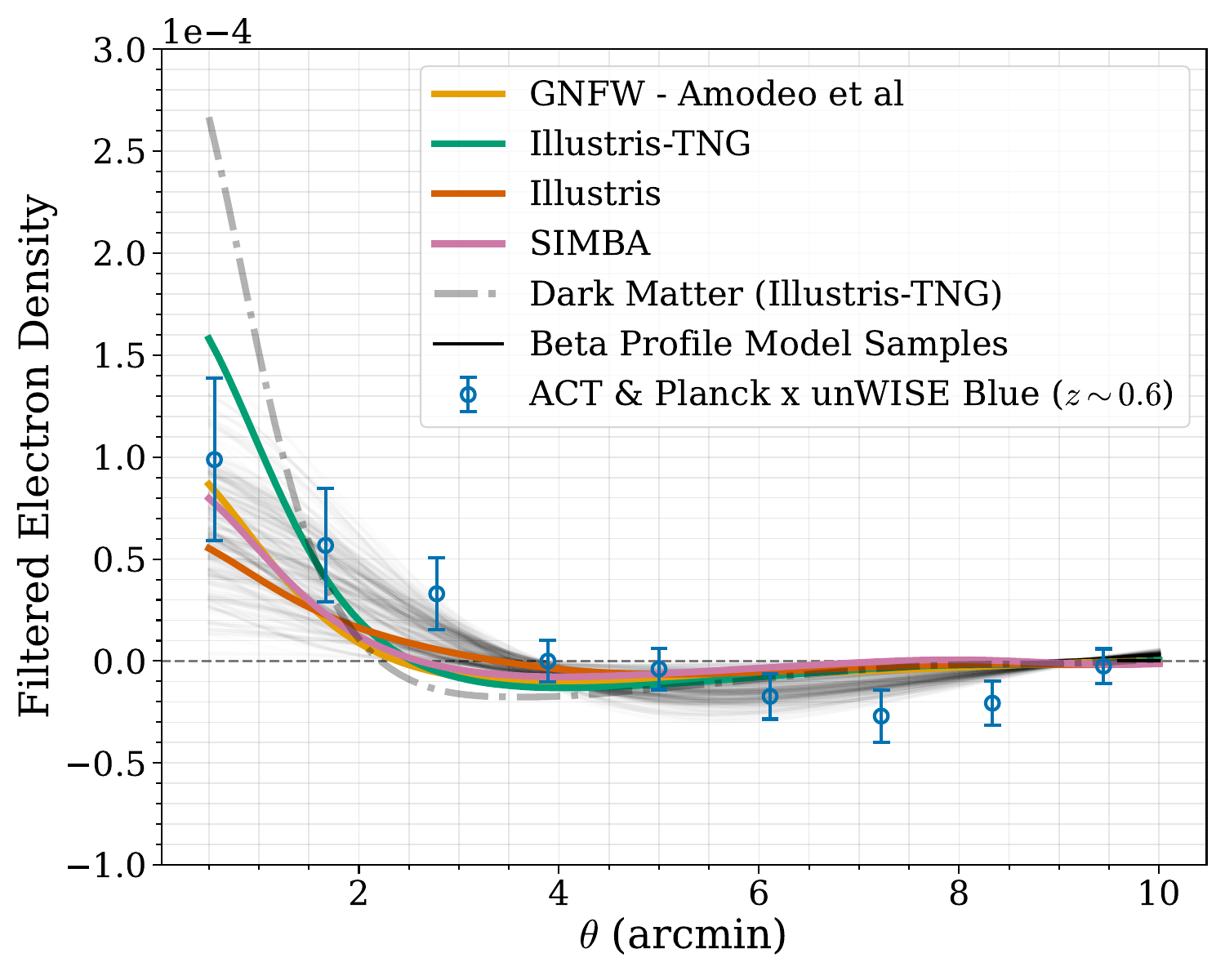} 
 \caption{
 Qualitative comparison of the measured profile to several theoretical models of the electron distribution. We compare the blue sample measurements (blue points, with $1\sigma$ errors) to the Illustris {(red)}, Illustris-TNG {(green)} and the SIMBA {(purple)} simulations, and the best-fit profile from the kSZ analysis of BOSS galaxies\cite{Amodeo_2021} {(orange)}. 
 The dark matter profile from Illustris-TNG is shown in dot-dashed grey, re-normalized to match the integrated electron density.
  The theoretical profiles have all been filtered in a manner equivalent to the data.
  The many, faint, black lines are posterior samples from the fit to the beta model. Whilst we do not detect this effect with ACT, these simulation predictions are suggestive that future, higher precision measurements will robustly measure this effect.
 \label{fig:dataVsTheory}}
\end{figure}

\section{Conclusions}\label{sec:conclusion}

CMB anisotropy data and large galaxy surveys have reached the precision necessary to study the anisotropic screening effect. 
This paper, along with Ref.~\cite{Schutt_2023}, introduces a new ``sign estimator'' robust to all other foregrounds and presents constraints on this effect. We find an upper limit on the amount of screening which translate into an integrated optical depth bound of $3.3\times 10^{-2}$ arcmin$^2$ at 95\% confidence.

The low signal-to-noise of this first measurement means that we are not able to meaningfully constrain the gas properties. However, it is still instructive to compare our constraint to past work and simulations.  In \cref{fig:dataVsTheory} we compare our optical depth measurements to a previous gas measurement from Ref.~\cite{Amodeo_2021}, shown in orange. Specifically we plot their best-fit generalized Navarro Frenk White (GNFW)\cite{Navarro_1996} profile. We also contrast our measurements with three state-of-the-art cosmological hydrodynamical simulations of galaxy formation: Illustris \cite{Nelson_2015}, IllustrisTNG \cite{Nelson_2019} and SIMBA\cite{Dave_2019}; and a prediction from the expected signal if the gas traced the dark matter.
For each simulation we follow the method of \cite{Thiele_2022,Lee_2022} to compute the profiles. 
We use the \textit{unWISE} properties from Ref. \cite{Kusiak_2022} to perform the comparisons, with the mean redshifts and halo masses listed in \cref{tab:unWISE_prop}. 
For the simulations, we use the nearest redshift snapshot and a mass bin of $1.5\times10^{13} M_\odot/h\leq M_\mathrm{halo}<2.4\times10^{13} M_\odot/h$. 
We do not attempt to reproduce the \textit{unWISE} sample selections, nor the impact of satellite galaxies.
\cref{fig:dataVsTheory} shows that our measurements are at the level of past measurements and the simulation predictions. Thus, with improved sensitivity screening measurements will be able to detect this effect and provide independent validations of kSZ measurements, such as those presented in \citep{Schaan_2021,Hadzhiyska_2024,Guachalla_2025}. 
This in turn may help resolve one of the largest uncertainties in interpreting gravitational lensing by galaxies as surveyed by Rubin, LSST, Euclid and Roman -- how matter is distributed in galaxies.

This approach complements existing CMB probes of the gas distribution \cite{Schaan_2016,Schaan_2021,Mallaby-Kay_2023,Kusiak_2021,Guachalla_2025,Hadzhiyska_2024}.
Unlike tSZ (resp. kSZ), it is unaffected by the gas temperature (resp. peculiar velocity).
Interestingly, measuring the anisotropic screening does not require any individual galaxy redshift information.
Together these three measurements can constrain  the full gas thermodynamics around galaxies (density, temperature, pressure and bulk motion).

\acknowledgements
We are very grateful to Leander Thiele for providing the code used to measure the gas profiles in the Illustris, Illustris-TNG and SIMBA simulations; and to Boryana Hadzhiyska and Noah Sailer for pointing out the lensing bias that was missed in the first version of this analysis.
Support for ACT was through the U.S.~National Science Foundation through awards AST-0408698, AST-0965625, and AST-1440226 for the ACT project, as well as awards PHY-0355328, PHY-0855887 and PHY-1214379. Funding was also provided by Princeton University, the University of Pennsylvania, and a Canada Foundation for Innovation (CFI) award to UBC. ACT operated in the Parque Astron\'omico Atacama in northern Chile under the auspices of the Agencia Nacional de Investigaci\'on y Desarrollo (ANID). The development of multichroic detectors and lenses was supported by NASA grants NNX13AE56G and NNX14AB58G. Detector research at NIST was supported by the NIST Innovations in Measurement Science program.  Computing for ACT was performed using the Princeton Research Computing resources at Princeton University, the National Energy Research Scientific Computing Center (NERSC),  and the Niagara supercomputer at the SciNet HPC Consortium. SciNet is funded by the CFI under the auspices of Compute Canada, the Government of Ontario, the Ontario Research Fund–Research Excellence, and the University of Toronto. We thank the Republic of Chile for hosting ACT in the northern Atacama, and the local indigenous Licanantay communities whom we follow in observing and learning from the night sky.

JCH acknowledges support from NSF grant AST-2108536, NASA grants 21-ATP21-0129 and 22-ADAP22-0145, the Sloan Foundation, and the Simons Foundation.
KM and MHi acknowledge support from the National Research Foundation of South Africa.
AK acknowledge support from NSF grant AST-2108536 
SN acknowledges support from a grant from the Simons Foundation (CCA 918271, PBL).
GAM is part of the Fermi Research Alliance, LLC under Contract No. DE-AC02-07CH11359 with the U.S. Department of Energy, Office of Science, Office of High Energy Physics
The Flatiron Institute is supported by the Simons Foundation.
CS and CV acknowledges support from the Agencia Nacional de Investigaci\'on y Desarrollo (ANID) through BASAL project FB210003.
MM acknowledges support from NSF grants AST-2307727, AST-2153201 and NASA grant 21-ATP21-0145.
NS acknowledges support from DOE award number DE-SC0020441.
ZA and JD acknowledge support from NSF grant AST-2108126.
ADH acknowledges support from the Sutton Family Chair in Science,
Christianity and Cultures, from the Faculty of Arts and Science,
University of Toronto, and from the Natural Sciences and Engineering
Research Council of Canada (NSERC) [RGPIN-2023-05014, DGECR-2023-
00180].
This material is based upon work supported by the National Science Foundation Graduate Research Fellowship under Grant No. DGE-2146755.
TS thanks the LSSTC Data Science Fellowship Program, which is funded by LSSTC, NSF Cybertraining Grant \#1829740, the Brinson Foundation, and the Moore Foundation; their participation in the program has benefited this work.

Data Access:
All the ACT data can be accessed on \url{https://lambda.gsfc.nasa.gov/product/act/actadv_prod_table.html|
}, the \textit{Planck} data are available at \url{https://pla.esac.esa.int/#home} and the unWISE data are available \url{https://catalog.unwise.me/}. 

\bibliographystyle{prsty.bst}
\bibliography{patchy,Planck_bib}

\end{document}